\documentclass[a4paper,11pt]{article}
\pdfoutput=1
\usepackage{jcappub}
\usepackage{caption}
\usepackage{hyperref}
\usepackage{amsmath}
\usepackage{mathtools}
\usepackage{tablefootnote}
\usepackage{soul}
\usepackage{ulem}
\usepackage{cancel}
\usepackage{color}
\usepackage{verbatim}
\usepackage{float}
\usepackage{graphicx,color}
\usepackage{subcaption}
\usepackage[usenames,dvipsnames,svgnames,table]{xcolor}
\usepackage{movie15}

\newcommand{\LCDM}{$\Lambda$CDM}
\newcommand{\thetade}{\theta_{\rm de}}
\newcommand{\deltade}{\delta_{\rm de}}
\newcommand{\thetab}{\theta_{\rm b}}
\newcommand{\deltab}{\delta_{\rm b}}
\newcommand{\cseff}{c_{\rm eff}}
\newcommand{\cs}{c_{\rm s}}
\newcommand{\dd}{{\rm d}}
\newcommand{\Sb}{\Gamma}
\newcommand{\mH}{{\mathcal H}}

\newcommand{\be}{\begin{equation}}
\newcommand{\ee}{\end{equation}}
\newcommand{\bea}{\begin{eqnarray}}
\newcommand{\eea}{\end{eqnarray}}

\title{On cosmological signatures of \\
baryons-dark energy elastic couplings}

\author[a]{Jose Beltr\'an Jim\'enez,}
\emailAdd{jose.beltran@usal.es}

\author[a]{Dario Bettoni,}
\emailAdd{bettoni@usal.es}

\author[a]{David Figueruelo,}
\emailAdd{davidfiguer@usal.es}

\author[a]{and Florencia A. Teppa Pannia}
\emailAdd{f.a.teppa.pannia@usal.es}

\affiliation[a]{Departamento de F{\'i}sica Fundamental and IUFFyM, Universidad de Salamanca, E-37008 Salamanca, Spain.}

\abstract{
We consider a scenario where dark energy and baryons are dynamically coupled without any energy transfer. In this scenario, the background cosmology is unaffected and, at the perturbations level, the coupling only appears through the corresponding Euler equations of dark energy and baryons. We then explore some phenomenological consequences of this scenario and their signatures in several cosmological observables. In particular, we show its ability to suppress the growth of cosmic structures. We also constrain the parameters of the model with cosmological data and show that an interaction of dark energy with baryons on cosmological scales is mildly favoured.}
 
\newpage
\begin{document}
\maketitle

\section{Introduction}
\label{sec:introduction}

Two decades after the discovery of the accelerated expansion of the Universe \cite{Perlmutter:1998np,Riess:1998cb}, we still lack a solid theoretical understanding of the underlying mechanism. The cosmological constant of General Relativity provides an efficient explanation able to account for a wealth of observations over a broad range of scales and times. Together with a cold dark matter component (DM), this so-called \LCDM~model succeeds in explaining data from the Cosmic Microwave Background (CMB) anisotropies \cite{Akrami:2018vks,Aghanim:2018eyx}, Large Scales Structures (LSS) \cite{Tegmark:2003ud}, Baryon Acoustic Oscillations (BAO) \cite{Blake:2011wn} or Type Ia supernovae (SNeIa) \cite{Conley:2011ku} and is nowadays firmly established as the standard model of cosmology. 

However, this picture suffers from the long-standing theoretical cosmological constant problem \cite{Weinberg:1988cp,Martin:2012bt} (with the related coincidence problem \cite{Zlatev:1998tr,Velten:2014nra}). This has impulsed a large effort in finding alternative scenarios to address the accelerated expansion where the role of the cosmological constant is replaced by some dynamical degree of freedom, dubbed dark energy (DE) \cite{Huterer:2017buf,Brax:2017idh}  (see also \cite{Clifton:2011jh,Joyce:2014kja}). On top of this theoretical shortcoming, some tensions in the data have appeared in recent years that are challenging the \LCDM~picture. In particular, the  inferred value of $H_0$
from CMB \cite{Aghanim:2018eyx} is in tension with the one measured by local experiments \cite{Riess:2019cxk,Wong:2019kwg,Birrer:2018vtm}. Another source of discomfort among observations is represented by the amplitude of matter fluctuations, conventionally encoded in the parameter $\sigma_8$, when measured locally \cite{Allen:2002eu} or at CMB \cite{Aghanim:2018eyx}.
Although unaccounted systematics could relax these tensions, this might be a hint towards new physics beyond \LCDM~ and pursuing this possibility is worth. Furthermore, the largely unknown properties of the dark sector further motivates exploring this possibility and ultimately only through observations can they be unveiled. 

Motivated by this emerging picture and the existing tensions, in this work we aim at exploring the intriguing, but feasible, possibility that DE could have appreciable interactions with ordinary baryonic matter at cosmological scales. We will focus here on a class of couplings leading to an elastic interaction such that there is no energy transfer between the two species, at least at first order in a perturbative expansion. Hence, no modifications to the cosmological background dynamics can be produced by virtue of the Cosmological Principle dictating that both fluids should have the same large scale rest frame. Though this represents a very minimal modification to the standard evolution, we will show in this work that it is sufficient to obtain interesting phenomenological consequences that can be tested by observations.
In this scenario, only the Euler equations of the coupled fluids are modified so non-trivial effects appear as soon as peculiar velocities start to grow due to the infall into the gravitational wells produced by means of the Jeans instability. This implies that, provided the interaction is not washed out by the cosmic expansion, it will start to be effective in the late-time Universe and at sub-horizon scales, affecting the matter clustering and hence modifying the amplitude of matter fluctuations. The interaction with the DE component can prevent the growth of matter fluctuations, thus potentially alleviating the $\sigma_8$ tension. It is important to notice, however, that the very definition of the interaction prevents any effect on the value of $H_0$. 
The implementation of the coupling closely follows that of \cite{Asghari:2019qld} between DM and DE where such an interaction is introduced as a modification to the fluid equations proportional to the relative velocities of the two components. Of course, there is no motivation for the coupling to occur only between a specific subset of the matter content of the Universe. However, our main motivation is to check first the consequences of such coupling between baryons and DE at the cosmological level as neatly as possible in order to clearly discern the genuine effects of the interaction without obscuring them. Enlarging the number of coupled species would produce a degeneration among the effects due to the various interactions, hence making it more cumbersome to understand the role of the baryonic one and, ultimately, its viability. Hence, the present work should be considered as a complementary investigation to the one carried out in \cite{Asghari:2019qld}. 

The described couplings that we will consider in this work resemble the interactions with pure momentum exchange that have also been explored in the context of DM-DE interactions. For example in \cite{Pourtsidou:2013nha,Skordis:2015yra} the conditions to have such interaction are given for scalar field-fluid system and their cosmological signatures have been studied in \cite{Pourtsidou:2016ico}. In particular, it was shown how these interactions can alleviate the aforementioned tensions \cite{Chamings:2019kcl}. Also, in \cite{Simpson:2010vh} it was suggested the possibility that DE and DM could interact via Thomson scattering and its consequences were subsequently explored in \cite{Baldi:2014ica,Baldi:2016zom,Kumar:2017bpv}.

Already in \cite{Simpson:2010vh} it was suggested that DE could present an elastic Thomson-like scattering with baryons and this idea has been pursued further in \cite{Vagnozzi:2019kvw}, where the authors showed the poor prospects to detect such couplings with cosmological observations. In this work, we will continue the study of possible interactions between baryons and DE from a different perspective. Although our approach is perhaps less theoretically motivated, it is more flexible and allows for a richer phenomenology. In particular, we will show how certain elastic interactions can in fact give non-negligible observational signatures. Exploring different parameterisations is of fundamental importance especially because, in the lack of a solid theoretical guiding principle, they can lead to draw different conclusions. The crucial difference of our scenario with respect to \cite{Vagnozzi:2019kvw} is the time dependence of the coupling and, as we will show, this can lead to significantly different results. The main reason is that assuming that baryons and DE scatter off each other as a pure Thomson process implies a coupling that decreases with the cosmic evolution, hence suppressing the interaction exactly in the regime where the velocities become relevant.
As we will explain in detail, our scenario features an effective coupling constant that grows in time and this will be  at the origin of the observational signatures that we will find. As we will argue below, the non-trivial DE background, that can be considered as a cosmologically evolving condensate, suggests that a Thomson-like scattering between DE and baryons could be mediated by an effective coupling that inherits the background time-dependence. Hence, there is also some theoretical motivation to go beyond the pure constant coupling case considered in \cite{Vagnozzi:2019kvw}. Of course, this phenomenological modelisation should be ideally embedded into a more fundamental framework where such interaction can be deduced from first principles. However, since our aim is to investigate the cosmological effects, we will not be concerned with these issues in the present work.

The paper is organised as follows. In Section \ref{sec:model} we will introduce the model and its main properties. In Section \ref{sec:analytical} we will provide  analytical solution to the equation in some detail to get an intuition of the relevant scales at play and to help with  the interpretation of the numerical results. These will be presented in Section \ref{sec:numerics}. In Section \ref{sec:fit}, we use the data available to constraint the cosmological and model parameters. Finally, in Section \ref{sec:conclusions} we draw our conclusions.

\section{The baryons-dark energy interacting model}
\label{sec:model}

In this Section we will introduce the phenomenological model that we will use in our subsequent analysis. As we have mentioned in the Introduction, we are interested in a class of models in which DE couples to baryons via an elastic interaction so there is no transfer of energy. The elastic nature of the interaction must be interpreted as the leading order effect in a perturbative expansion, while at some order in perturbation theory a transfer of energy will also appear. For the cosmological scenario we are interested in this work, it will suffice to assume that the interaction is elastic at least up to first order in cosmological perturbations. In particular, the very definition of the interaction ensures that the background cosmology remains completely unaffected. Thus, if we assume that the Universe components can be described in terms of perfect fluids, the total energy-momentum tensor is 
\begin{equation}
T^{\mu\nu}=\sum_i\Big[(\rho_{i} +p_{i})u_i^\mu u_i^{\nu} +g^{\mu\nu}\, p_{i}\Big]\;,
\end{equation}
where $\rho_{i}$, $p_{i}$ and $u_i^\mu$ are the energy-density, pressure and the 4-velocity of the $i$-th component and the sum runs over all the components of the Universe. The Cosmological Principle dictates that all the components share the same large scale rest frame so they have the same zeroth order velocity and the background metric is described by the flat Friedman-Lema\^itre-Roberton-Walker line element
\be
\dd s^2=a^2(\tau)\Big(-\dd \tau^2+\dd \vec{x}^2\Big)\,.
\ee
The continuity equations are the usual ones
\begin{equation}
\rho_{i} ' + 3 \mathcal{H}\left(1 +w_{i}\right)\rho_{i}=0\,,
\end{equation}
with $w_{i}=\frac{p_{i}}{\rho_{i}}$ the equation of state parameter,
and the Friedmann equation also remains the same
\begin{equation}
\mathcal H^2 = \frac{8 \pi G a^2}{3 }\sum_i \rho_{i}\,,
\end{equation}
with $\mH=a'/a$ the Hubble expansion rate. The interesting sector of the model under consideration concerns the linear perturbations and, in particular, the Euler equations where the interaction will enter. We will work in Newtonian gauge (see Appendix \ref{app:synch} for the equations in synchronous gauge) so the perturbed line element for scalar modes reads
\begin{equation}
\dd s^2=a^2(\tau)^2 \Big[-(1+2 \Psi)\dd \tau^2 +(1-2 \Phi)\dd \vec{x}^2\Big] \;,
\end{equation}
where $\Phi$ and $\Psi$ are the gravitational potentials. We assume that the first order perturbations of the energy-momentum tensor also take the form of a perfect fluid. In particular, this guarantees the absence of anisotropic stresses (at least in the relevant scales for us) and, given that Einstein equations are not affected by the interaction, the slip parameter $\gamma\equiv \Phi/\Psi$ is 1 so that $\Phi=\Psi$.

For the matter sector equations, since we only introduce a coupling between DE and baryons, the equations for photons, neutrinos and DM do not change. The interacting sector is however governed by the following system of coupled equations (using the definitions in~\cite{Ma:perturbations}):
\begin{eqnarray}
\label{eq:dotdeltab}
\deltab'&=&-\thetab + 3 \Phi' \;,\\
\label{eq:dotthetab}
\thetab'&=&-\mathcal{H} \thetab+ k^2 \Phi
+\Gamma_T(\theta_{\gamma}-\thetab)+\Gamma (\thetade-\thetab)\;,\label{eq:thetaB}\\
\deltade'&=&-3 \mathcal{H}(\cs^2-w)\deltade-(1+w)\left[1+9 \frac{\mathcal{H}^2}{k^2}\left(\cs^2-w\right)\right]\thetade
 +3(1+w)\Phi' \;,\\
\thetade'&=&(-1+3\cs^2) \mathcal{H}\thetade+k^2\Phi +\frac{k^2 \cs^2}{1+w}\deltade-\Gamma R(\thetade-\thetab)\;,
\label{eq:thetaDE}
\end{eqnarray}
where we have used the standard notations for the density contrast $\delta\equiv\frac{\delta \rho}{\rho}$ and for the Fourier space velocity perturbation  $\theta\equiv i \vec{k}\cdot\vec{v}$. We also assume a constant DE equation of state parameter, i.e. $w'=0$. As advertised above, the continuity equations are not modified and only the Euler equations receive corrections due to the interactions. For the baryons we have the standard term to describe the Thomson scattering with photons $\Gamma_T\equiv\frac{4\rho_{\gamma}}{3 \rho_{b}}a n_e \sigma_T$ with $n_e$ the abundance of free electrons and $\sigma_T$ the Thomson scattering cross Section. Besides this coupling, that is only relevant when there is a non-negligible fraction of free charges before recombination and during reionisation, we have introduced the novel interaction term driven by $\Gamma$ in the baryons Euler equation and $R\Gamma$ in the DE sector, with
\be
R  \equiv  \label{eq:Rcoupling} \frac{\rho_{\rm b}}{(1+w)\rho_{\rm de}} = \frac{\Omega_{\rm b}}{(1+w)\Omega_{\rm de}}a^{3w}\;,
\ee
the baryon-to-dark energy ratio. Notice that this ratio is $R\gg1$ for most of the cosmic history. Following the analogy with the Thomson scattering, we can parameterise the interaction rate as
\begin{eqnarray}
\label{eq:Gcoupling}
\Gamma &\equiv& \bar{\beta}\frac{a}{\rho_{\rm b}}\;,
\end{eqnarray}
where $\bar{\beta}$ plays the role of the effective coupling constant. It is convenient to normalise the coupling constant as follows
\begin{equation}
\beta=\frac{8  \pi G}{3 H_0^3} \bar{\beta}.
\end{equation}
With this normalisation, the natural scale associated to the interaction is $\Lambda_\beta\sim (M_{\rm Pl}^2 H_0^3)^{1/5}\sim 10^{-9}$ eV, which will be confirmed by our observational fits favouring values of $\beta=\mathcal{O}(1 - 10)$. This scale is small even as compared to the meV scale of DE. Let us note however that this does not necessarily reflect the relevant scale of the underlying model. In the case of a Thomson-like scattering this parameter would be proportional to the cross Section and the abundance of interacting particles. However, let us emphasise that we do not necessarily adopt this approach and will take $\beta$ as a general coupling, that could be time and/or scale dependent, giving an effective description for the linear perturbations without specifying the underlying microphysical model. This viewpoint can be formalised at the covariant level by assuming that the (non-)conservation equations of the baryons-DE interacting sector on the relevant scales take the form
\begin{equation}
\nabla_\mu T^{\mu \nu}_{\rm b}=Q^\nu\;,\quad\quad 
 \nabla_\mu T^{\mu \nu}_{\rm de}=-Q^\nu\;,
\end{equation}
with an interaction $Q^\mu$ that only affects the momentum conservation equation up to first order. A simple choice is to tie the interaction to the relative motion of the components so that it is proportional to their relative velocity:
\begin{equation}\label{eq:interaction_covariant}
 Q^{\mu}=
 \bar{\beta}\left( u^\mu_{\rm de}-u^\mu_{\rm b} \right)\;.
\end{equation}
This clearly reproduces the desired feature of only affecting the Euler equations by virtue of the assumed common large scale rest-frame of all the Universe components. This phenomenological coupling has been considered in the context of DM-DE interactions in \cite{Asghari:2019qld}. In the same context, there are other scenarios where similar interactions have been considered. In \cite{Simpson:2010vh}, for example, this coupling arises by assuming a Thomson scattering with DE. This Thomson scattering could be understood as an interaction of DM/baryons with the DE phonons,\footnote{This interpretation was not explicitly mentioned in \cite{Simpson:2010vh} and the subsequent studies \cite{Baldi:2014ica,Baldi:2016zom,Vagnozzi:2019kvw} but it is the one we find most appealing.} interpreted as the fluctuations of the DE condensate over its cosmologically evolving background. In this sense, the Thomson scattering would be different from the one taking place before recombination between photons and baryons where the interacting particles conform a thermal distribution with trivial background field values. Thus, one could expect the coupling constant in the Thomson scattering with DE to depend on the background value of DE condensate, thus inducing a cosmological evolution. This scenario is the one that was considered in more detail in \cite{Vagnozzi:2019kvw} to explore possible interactions between baryons and DE, but a constant coupling constant was utilised there. They concluded that effects on cosmological scales are too small to be observable or to have a relevant impact on them. This can be traced to the fact that the interaction rate decreases with the expansion so at late times is very small and at high redshift when it can be relevant, the DE component is negligible. However, allowing for a more general scenario where the coupling constant varies over cosmological scales permit non-negligible effects, for instance if $\Gamma$ grows throughout the Universe expansion. It is straightforward to see that a constant $\beta$ leads to having $\Gamma\propto a^4$ so it grows substantially in time and can compensate for the small fraction of baryons at late times. One could motivate a constant $\beta$ at a very phenomenological level as it is the simplest case in view of \eqref{eq:interaction_covariant} or at a slightly more fundamental level by assuming that DE is described by a scalar field $\phi$ and the interaction is mediated by the gradient of the scalar field so that, for the relevant regime, we would have $\beta\propto \dot{\phi}$. If DE has an approximate shift symmetry, then it is expected to have $\phi\propto t$ for its background evolution so that $\beta\propto\dot{\phi}$ is approximately constant. Needless to say that this is not a rigorous argument but it gives support to having a constant $\beta$. We will show in this work that, unlike the findings in \cite{Vagnozzi:2019kvw}, it is possible to have detectable signatures of a baryons-DE elastic interaction when a general time evolution of the coupling is included. This will be the main focus of the subsequent Sections.

Finally, let us  briefly comment on the behaviour of the interaction on small scales. It is a very well known fact that any interaction between DE and baryonic matter is severely constrained by Solar System tests and laboratory experiments. Indeed, DE models that contain such coupling are usually required to come equipped with some mechanism that suppresses the coupling at small scales. In this respect, it is noteworthy to  observe that in the  model under consideration the interaction  \eqref{eq:Gcoupling} goes as $\Gamma \propto \rho_{b}^{-1}$ hence decreasing for increasing baryon density. It is tempting to extrapolate such linear theory behaviour to astrophysical or Solar System scales and conclude that our model is endowed with a natural screening mechanism that suppresses the interacting term on small scales, thus reconciling its predictions with local observations. Although appealing, we should be careful taking this path. In fact, the form of the interaction rests on few assumptions that are true at cosmological scales but that may be unreliable on smaller ones. For example, it assumes baryons can be well described as a perfect fluid, it ignores non-linear effects, etc.  Hence, we should be cautious when extending the present interaction outside the cosmological regime. It would be, nonetheless, interesting to investigate how to embed the presented model in a more fundamental or theoretically motivated framework that could allow to explore also astrophysical and Solar System scales. However, since in the present work we are interested in the cosmological analysis, we will not delve into these issues. The reader should bear in mind however that our description is purely phenomenological and applicable to cosmological scales.

\section{Analytical analysis}
\label{sec:analytical}
In this Section, we will provide some analytical insights on the coupled system  \eqref{eq:dotdeltab}--\eqref{eq:thetaDE} that will be useful to interpret the numerical results below as well as to identify the relevant scales in the problem. We will generically consider a background dominated either by radiation or by (dark) matter while, at perturbation level, we will assume that DM represents the dominant contribution to the gravitational potentials. 

\subsection{Evolution outside the horizon}

Let us start our analytical analysis by considering the evolution of super-Hubble modes which are directly connected with the primordial perturbations generated from inflation in the early Universe. As usual, it is expected that these modes inherit the adiabatic nature of the primordial perturbations which is crucial to set appropriate initial conditions for the perturbations. This will be guaranteed by ensuring that the homogeneous solutions of the perturbations equations in this regime decay sufficiently quickly so the primordial adiabatic mode dominates. Had we growing homogeneous solutions, setting the initial conditions would be more subtle. Since the interaction is proportional to the relative velocity and this is very small on large scales, we can expect the effects to be negligible outside the horizon. In the following we will explicitly show that this is indeed the case even when the interaction term dominates. The equations for the super-Hubble modes can be approximated by 
\begin{eqnarray}
\deltab'&=&-\thetab  \;,\\
\thetab'&=&-\mathcal{H} \thetab + \frac{4\rho_{\gamma}}{3 \rho_{b}}a n_e \sigma_T(\theta_{\gamma}-\thetab)+\Gamma (\thetade-\thetab)+ k^2 \Phi \;,\\
\label{eq:thetaBSH}
\deltade'&=&-3 \mathcal{H}\Big(\cs^2-w\Big)\deltade-9(1+w)\Big(\cs^2-w\Big)\frac{\mathcal{H}^2}{k^2}\thetade\;,\\
\thetade'&=&-\Big(1-3\cs^2\Big) \mathcal{H}\thetade+\frac{k^2 \cs^2}{1+w}\deltade-\Gamma R(\thetade-\thetab)+ k^2 \Phi \;.
\end{eqnarray}
At sufficiently high redshift the interaction rate is inefficient as compared to the Hubble expansion, $\mH\gg R\Gamma\gg\Gamma$, so we can neglect the interaction. In this case baryons evolve as usual. For the DE perturbations, we can consider the gravitational potential $\Phi$ as an external source. If we assume a power law expansion with $\mH=p/\tau$, as it would be for a single species dominated background, it is easy to see that the homogeneous solution has the form $\deltade=A\tau^{n-1}$ and $\thetade=B \tau^n$. If we insert this Ansatz in the homogeneous equations for DE we obtain
\bea
B_\pm&=&\frac{1-n_\pm-3p(\cs^2-w)}{9(\cs^2-w)(1+w)}\frac{k^2A_\pm^2}{p^2}\;,\\
n_\pm&=&\frac12\left[1 - p (1 - 3 w) \pm \sqrt{1 - 12 \cs^2 p (1 + p) + p (1 + 3 w) (2 + p + 3 p w)}\,\right]\;.
\eea
If we take $\cs^2=1$ and $w\simeq-1$, but $w\neq-1$ in order to avoid the singular case of a cosmological constant, we obtain that $n_\pm$ pick an imaginary part that leads to an oscillatory behaviour with an amplitude that decays with $\text{Re}(n_\pm)\simeq-3/2$ during radiation domination ($p=1$) and with $\text{Re}(n_\pm)\simeq-7/2$ during matter domination ($p=2$). The perturbations are then attracted to the inhomogeneous solution driven by $\Phi$. This is the usual super-Hubble evolution for DE perturbations and the explicit expressions can be found in e.g. \cite{Ballesteros:2010ks,Asghari:2019qld}.

In the presence of the interaction between baryons and DE,  the homogeneous solution is modified as the interaction terms become relevant in the Euler equations. This occurs first for DE because $R\gg1$ so we have a regime where $R\Gamma\gg\mH\gg\Gamma$. Under these circumstances, the equation for baryons remains oblivious to the interaction so it evolves as usual and we can consider it as an external source in the DE Euler equation. On the other hand,  using that $k^2\deltade\sim\mH\thetade\ll R\Gamma\thetade$ we can also neglect the Laplacian term originating from pressures in the Euler equation of DE. The peculiar velocity of DE then decouples and it is solely driven by the interaction that in turn gives $\thetade\propto e^{-\int R\Gamma\dd\tau}$. This clearly shows that it decays exponentially so its contribution to the DE continuity equation is negligible and the homogeneous solution for $\deltade$ also decays. Hence, also in this regime the evolution is attracted to the inhomogeneous solution of the equations that is now determined by both $k^2\Phi$ and $\Gamma R\thetab$. Since $\thetab$ can be obtained from the baryons Euler equation as $\thetab\simeq \frac{1}{p+1}k^2\Phi\tau$, we have that the relative importance of both terms is
\be
\frac{\Gamma R\thetab}{k^2\Phi}\simeq \frac{p}{p+1}\frac{\Gamma R}{\mH}\gg1\;,
\ee
so one might think that the baryons peculiar velocity drives the DE evolution. We have to note however that the pure adiabatic mode generates the same velocity perturbation for all the components, so even if $\Gamma R\gg1$, the large scale common rest frame makes the relative velocity $\thetade-\thetab$ very small and the interaction term is actually negligible for super-Hubble modes. This will be relevant in the sub-horizon regime. Something similar happens in the regime with $\Gamma\gg\mH$. In summary, the decaying nature of the homogeneous solutions in all the regimes shows that the super-Hubble evolution is driven by the inhomogeneous solution that is determined by the primordial spectrum. This can affect the DE evolution in two ways, however, that we will clarify below.

\subsection{Dark energy - baryons tight coupling approximation}
After clarifying the evolution of super-Hubble modes, let us turn to the more interesting regimes where the interaction gives noticeable effects. At low redshift and for sub-Hubble modes, the interaction rate $\Sb$ becomes very large so that DE and baryons will form a locked system similar to the photon-baryon fluid before recombination due to Thomson scattering. In this regime, there is an efficient dragging generated by the interaction that makes $\thetade\simeq\thetab$. For sufficiently small scales there are additional effects that will reduce the dragging efficiency of the interaction and baryons no longer follow the DE flow, as we will discuss in the next Sections. For the scales where the dragging is efficient we can obtain a decoupled equation for the evolution of the DE density. Firstly, we note that the momentum conservation for baryons can be written as
\be
\thetab=\thetade-\frac{1}{\Sb}\Big(\thetab'+\mH\thetab-k^2\Phi\Big)\;,
\ee
where we have neglected the contribution from Thomson scattering with photons\footnote{It is important to notice the competing effect of the Thomson scattering and the interactions with DE in the baryons Euler equation. Since before recombination the coupling to DE is negligible, the Thomson scattering provides the dominant dragging. At the epoch of reionisation however the coupling to DE can be relevant for sufficiently large $\beta$ and this could affect the evolution of the perturbations at that epoch. We will confirm that this can be the case from our full numerical treatment below (see Fig. \ref{fig:evolTERMSz}).}. As expected, in the regime with a large interaction $\Gamma\gg\mH$, the two velocities are approximately the same provided the gravitational term in the bracket is not too large $k^2\Phi\lesssim \mH\thetade$. The first order correction to the difference of the velocities is then given by
\be
\thetab^{(1)}=\thetade-\frac{1}{\Sb}\Big(\thetade'+\mH\thetade-k^2\Phi\Big)\;.
\ee
We can introduce this relation into the Euler equation for DE and combine it with the DE continuity equation to obtain the following second order differential equation governing the evolution of the DE density contrast:
\be
\deltade''+\Big[1-3\big(w-\cseff^2 R\big)\Big]\mH\deltade'+\Big(\cseff^2 k^2+m_{\rm eff}^2\Big)\deltade=(1+w)\Big[-k^2\Phi+3(1-3\cseff^2)\mH\Phi'+3\Phi''\Big]\;,
\label{eq:dDEtight}
\ee
with
\be\label{eq:meff}
\cseff^2=\frac{c_{\rm s}^2}{1+R}\quad{\text {and}}\quad m_{\rm eff}^2=3(c_{\rm s}^2-w)\Big[(1-3\cseff^2)\mH^2+\mH'\Big]\;.
\ee
Since the RHS of \eqref{eq:dDEtight} only depends on the gravitational potential and this is essentially determined by the DM clustering, we can treat it as an external source for the DE density contrast to a good approximation. Furthermore, as we have mentioned above, we can take $\Phi$ to be constant so we will neglect its time derivatives (though they can be relevant at very late times when DE dominates). On the other hand, since we have that $R=\frac{\Omega_{B}}{(1+w)\Omega_{DE}} a^{3w}\gg 1$ throughout most of the Universe evolution, we can further simplify the friction coefficient so that the equation can be expressed in the approximate form
\be
\deltade''+\Big[1-3\big(w-\cs^2\big)\Big]\mH\deltade'+\left(\frac{\cs^2}{R} k^2+m_{\rm eff}^2\right)\deltade=-(1+w)k^2\Phi\;.
\label{eq:dDEtight2}
\ee
This equation shows that, as long as the tight coupling approximation holds and the gravitational wells are not too deep, the evolution for the DE density contrast will correspond to a damped oscillator with a constant driving gravitational force. As we said above, the DE component only contributes negligibly to the Poisson equation that determines $\Phi$, so the RHS will only contribute to the inhomogeneous part of the solution. The homogeneous solution can be found by using the WKB approximation in the sub-horizon regime. Let us notice, however, that the sub-horizon approximation presents two regimes depending on the relative hierarchy between the effective mass $m_{\rm eff}^2$ and the sound horizon $\cs^2 k^2_{\rm s}\simeq R\mH^2$. The two independent WKB solutions to the equation are given by
\be
\deltade^{\pm}\propto \frac{1}{\sqrt{a^{1-3(w-\cs^2)}\sqrt{\frac{\cs^2}{R} k^2+m_{\rm eff}^2}}}e^{\pm i\displaystyle\int\sqrt{\frac{\cs^2}{R} k^2+m_{\rm eff}^2}\dd\tau}\;,
\ee
which, for modes inside the sound horizon, simplifies to
\be
\deltade^{\pm}\propto \sqrt{\frac{\sqrt{R}}{a^{1-3(w-\cs^2)}}}e^{\pm i\displaystyle\int\frac{\cs k}{\sqrt{R}} \dd\tau}\propto a^{(9w-8)/4}e^{\pm \frac{i}{\sqrt{R_0}}\displaystyle\int a^{3/2}k \dd\tau}\;,
\ee
where $R_0=\frac{\Omega_{B}}{(1+w)\Omega_{DE}}$ and we have used that $\cs^2=1$. The particular solution can be found with the general formula
\be
\deltade^p=-(1+w)k^2\Phi\left[\deltade^+\int\frac{\deltade^-}{\det W}\dd \tau-\deltade^-\int\frac{\deltade^+}{\det W}\dd \tau\right]\;,
\ee
with $\det W=(\deltade^+)'\deltade^--\deltade^-(\deltade^+)'$ the Wronskian. In the WKB approximation that we are considering, the Wronskian can be approximated as $\det W\simeq a^{(9w-8)/4}2i\sqrt{\frac{\cs^2 k^2}{R}+m_{\rm eff}^2}$ and the particular solution can be written as
\be
\deltade^p\simeq-(1+w)\frac{k^2}{\frac{\cs^2k^2}{R}+m_{\rm eff}^2}\Phi\;,
\ee
where we have used that $k\tau\gg1$ so $R$ and $a$ evolve very slowly and can be taken outside the integrals. Deep inside the horizon, this particular solution reduces to
\be
\deltade^p\simeq-\frac{1+w}{\cs^2}R\Phi\;.
\ee
We then see that the density contrast consists of an oscillating piece whose amplitude decays as $\propto a^{-17/4}$ on top of the above particular piece whose amplitude is determined by the gravitational potential and decays as $R\propto a^{-3}$. Since the amplitude of the oscillations decays much faster, the DE density contrast will eventually be dominated by the particular solution. We can thus expect that the density contrast oscillates when it enters the horizon (and the interaction dominates over the Hubble expansion) with a decaying amplitude and eventually the particular solution takes over. We will confirm this with the numerical solutions. However, for sufficiently small scales  other important effects appear that we analyse in the following.

\subsection{Baryonic pull}
As we have discussed above, there is a period of time over which the interaction term is still negligible for baryons, but it drives the evolution of DE, as can be deduced from the definition of the couplings \eqref{eq:Rcoupling} and \eqref{eq:Gcoupling}. In that regime, the peculiar velocities of baryons dominate over those of DE so we can approximate $\Sb R(\thetade-\thetab)\simeq -\Sb R\thetab$ and the DE equations can be combined to give 
\be
\deltade''+\Big[1-3\big(w-\cs^2\big)\Big]\mH\deltade'+\left(\cs^2 k^2+m_{\rm eff}^2\right)\deltade=\Sb R\thetab\;,
\ee
with $m_{\rm eff}$ given in \eqref{eq:meff} and where the RHS can be considered an external source. In the matter dominated epoch, the solution is dominated by the inhomogeneous part that gives
\be
\deltade\simeq \frac{\Sb R}{\cs^2 k^2}\thetab\;,
\ee
where $\thetab$ follows the usual evolution of a \LCDM \ Universe and we have used that $\cs^2k^2\gg m_{\rm eff}^2$. We can insert this solution into the continuity equation of DE to obtain its velocity perturbation 
\be
-(1+w)\thetade\simeq\deltade'+3\mH(\cs^2-w)\deltade,
\ee
that allows to obtain
\be
-(1+w)\thetade\sim \frac{\Gamma R \mH}{\cs^2 k^2}\thetab
\ee
that shows the validity of this regime for scales $\cs k\gg \sqrt{\Gamma R\mH}$.

\subsection{Gravitational pull}
On small scales there are two competing effects, namely:  the dragging generated by the baryons-DE interaction and the gravitational pull. These two effects give rise to two regimes depending on which of the two dominates. For large enough scales, the gravitational pull is weaker than the dragging so that while the inertia carried by the baryons tends to make the system fall into the gravitational wells, the DE pressure gradients opposes to this collapse and the net effect is the acoustic oscillations discussed above. However, on sufficiently small scales, the gravitational wells are deep enough so that the dragging ejected by the DE pressure gradients is not sufficient to prevent the collapse. The acoustic oscillations cease and the baryons fall into the gravitational potentials, but more slowly than usual due to the interaction with DE. In this regime, the velocity of baryons is much larger than that of DE because the latter is still directly subject to its own gradient pressures. Thus, we can neglect $\thetade$ against $\thetab$ and the Euler equation for baryons simplifies to
\be
\thetab'+\Gamma\thetab=k^2\Phi\;,
\label{eq:thetabeq}
\ee
where we have also used that $\Gamma\gg\mH$. We can introduce the convenient time variable $\dd x=\Gamma\dd \tau$ that gives a measurement of time in terms of the interaction rate, so that  we can rewrite the equation as
\be
\frac{\dd\thetab}{\dd x}+ \thetab=\frac{k^2\Phi}{\Gamma}\;,
\ee
whose general solution can be easily found to be
\be
\thetab=C_1e^{-x}+e^{-x}\int e^x\frac{k^2\Phi}{\Gamma}\dd x\;,
\label{eq:solthetab}
\ee
with $C_1$ an integration constant. We will assume now that we are will within the matter domination epoch so that the gravitational potential is constant and $a=\tau^2$. The interaction rate is then $\Gamma=\Gamma_0\tau^8$ so that $x=\frac{1}{9}\Gamma_0 \tau^9=\frac19\Gamma_0 a^{9/2}\gg1$, which simply reflects that we are in a regime where the interaction time-scale is much shorter than a Hubble time. Thus, the first term in \eqref{eq:solthetab} is exponentially suppressed and the baryons peculiar velocities are driven by the gravitational infall, as expected. The integral in \eqref{eq:solthetab} can be obtained in terms of incomplete Gamma functions, but we will not need it here. Instead, we can notice that, since $x\gg1$, the exponential inside the integral varies much faster than $\Gamma$, that only varies as the power law $\Gamma\propto x^{8/9}$. This permits to take it out of the integral so that the baryons velocity is simply
\be
\thetab\simeq\frac{k^2\Phi}{\Gamma}=\frac{k^2\Phi}{\Gamma_0} a^{-4}\;.
\ee
This result could have been obtained directly from \eqref{eq:thetabeq} by noticing that $\thetab'\simeq\mH\thetab\ll\Gamma\thetab$. The density contrast of baryons can be computed straightforwardly from the baryons continuity equation
\be
\deltab'\simeq -\thetab\Rightarrow\deltab\simeq\deltab^0+\frac{k^2\Phi}{7\Gamma_0}a^{-7/2}\;,
\ee
where $\deltab^0$ is the integration constant that accounts for the initial amplitude. We see that the rapid decay of $\thetab$ makes the density contrast be insensitive to the increasing depth of the potential wells and its amplitude remains constant. This behaviour is in high contrast to the usual growth $\propto a$ in the absence of the interaction with DE and is the origin of the suppression in the matter power spectrum on those scales.

We can now turn to the evolution of the DE perturbations. In the considered regime, the continuity and Euler equations can be approximated to
\bea
\deltade'&=&-3\mH(\cs^2-w)\deltade-(1+w)\thetade\;,\\
\thetade'&=&\frac{\cs^2k^2}{1+w}\deltade+(1+R)k^2\Phi\;,
\eea
where we have neglected the time-dependence of $\Phi$ and used the solution for the baryons velocity $R\Gamma(\thetade-\thetab)\simeq-Rk^2\Phi$. Furthermore, we have used that $\mH\thetade$ is negligible against $k^2\Phi=\Gamma\thetab$ because $\Gamma\gg\mH$ and $\thetab\gg\thetade$. We can combine the above equations to obtain the following equation that governs the DE contrast evolution:
\be
\deltade''+3(\cs^2-w)\mH\deltade'+\cs^2k^2\deltade=-(1+w)(1+R)k^2\Phi\;,
\label{eq:deltappsmall}
\ee
where we have used again that $\cs^2k^2\gg\mH^2$. This equation can be solved exactly, but it will be more instructive to obtain the WKB approximate solutions, as it is the relevant part for the regime under consideration. The homogeneous solutions are
\be
\deltade^{\pm}=\frac{1}{a^{3(\cs^2-w)/2}}e^{\pm ik\tau}\;.
\ee
We can follow the same procedure as before to obtain the particular solution
\be
\deltade^p\simeq-\frac{1+w}{\cs^2}(1+R)\Phi\;,
\ee
that can be directly obtained from \eqref{eq:deltappsmall} by dropping the time-derivatives, since the gravitational source only exhibits a mild time-dependence.

\section{Numerical analysis}
\label{sec:numerics}

In the previous Section we have obtained some analytical insights on the consequences of the baryons-DE interacting model. We will now explore its effects on several observables by a complete numerical analysis. This will allow us to confirm the findings of Section~\ref{sec:analytical} and gain a deeper understanding of the different regimes.

We have modified the publicly available codes CLASS~\cite{Julien:CLASS,Blas:CLASSaprox} and CAMB~\cite{Lewis:1999bs} for the computation of the evolution of linear perturbations. We have found no differences between the results obtained with the modifications of both codes. The background cosmology is not affected as we have explained and, therefore, we only need to modify the codes by adding the new interaction term in Euler equation for both DE and baryons.\footnote{The equation of the velocity perturbation of photons must be also modified in accordance with the original numerical scheme described in Section~5.7 of Ref.~\cite{Ma:perturbations} for CAMB and following the default scheme of CLASS described in Ref.~\cite{Blas:CLASSaprox}.} As a consequence, any deviation from the standard model is only due to the modification in the perturbation sector. We also explore in both codes possible effects due to the corresponding modifications for the Tight Coupling Approximation (TCA) and the Radiation Streaming Approximation (RSA) schemes described in Ref.~\cite{Doran2005,Blas:CLASSaprox}, consistent with baryon sector modifications. However, since the coupling term becomes relevant at low redshifts, where these approximations are not relevant, no differences w.r.t. the original schemes were found.

In the following results, we fix the cosmological parameters to $H_0=67.4$ km/s/Mpc, $\Omega_b h^2=0.0224$, $\Omega_{dm}h^2=0.120$, $w=-0.98$ and $c_s^2=1$. Our reference model is a $w$CDM rather than \LCDM \ to clearly isolate the effects of the interaction and to avoid degeneracies with, e.g., the DE equation of state. Notice that the interaction requires $w\neq-1$ since a pure cosmological constant does not have perturbations. Any other parameter is set to the default value of each code.

We organise this Section as follows. We first investigate the impact of the baryons-DE coupling in the matter power spectrum and in the CMB.  After that, in order to understand the produced new features, we show the evolution of the baryon density contrast and how the clustering is modified by plotting the values of $\sigma_8$ as a function of the coupling parameter $\beta$. Finally, as the interaction is determined by the relative velocity between the two fluids, we show how this gauge invariant quantity is modified by the interaction and interpret it in view of our results in Section~\ref{sec:analytical}.

\subsection{Different regimes}
\label{sec:regimes}
\begin{figure}[!t]
    \centering{
	\includegraphics[width=0.7\textwidth]{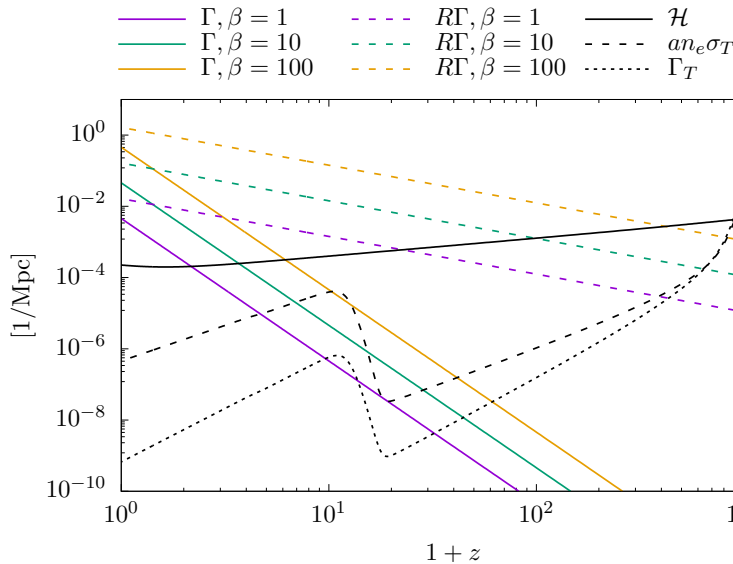}
	}
	\caption{Cosmological evolution of the interaction rates for different values of the parameter $\beta$. In all cases, the interaction becomes relevant for DE much earlier than for baryons. The Thomson scattering rates for baryons and photons are also included for comparison.}
	\label{fig:regimes}	
\end{figure}

In this Section, we provide the numerical counterpart of the analytical insights presented in Section~\ref{sec:analytical}. The cosmological evolution of the interaction functions $\Gamma$ and $R\Gamma$ for three different values of $\beta$ together with the Hubble function $\mathcal H$ is shown in Fig.~\ref{fig:regimes}. We also include, for comparison, the strength of the Thomson scattering between baryons and photons. 
Consistently with our previous analysis, at sufficiently high redshift the interaction can be neglected for both baryons and DE and the terms associated to the Thomson scattering evolve as usual. However, since $R\gg 1$, the DE interaction term becomes relevant earlier than the one entering the equation for baryons, hence confirming the hierarchy among the interaction terms and the Hubble rate discussed in Section \ref{sec:analytical}. 

\begin{figure}[!t]
	\centerline{
		\includegraphics[width=0.5\textwidth]{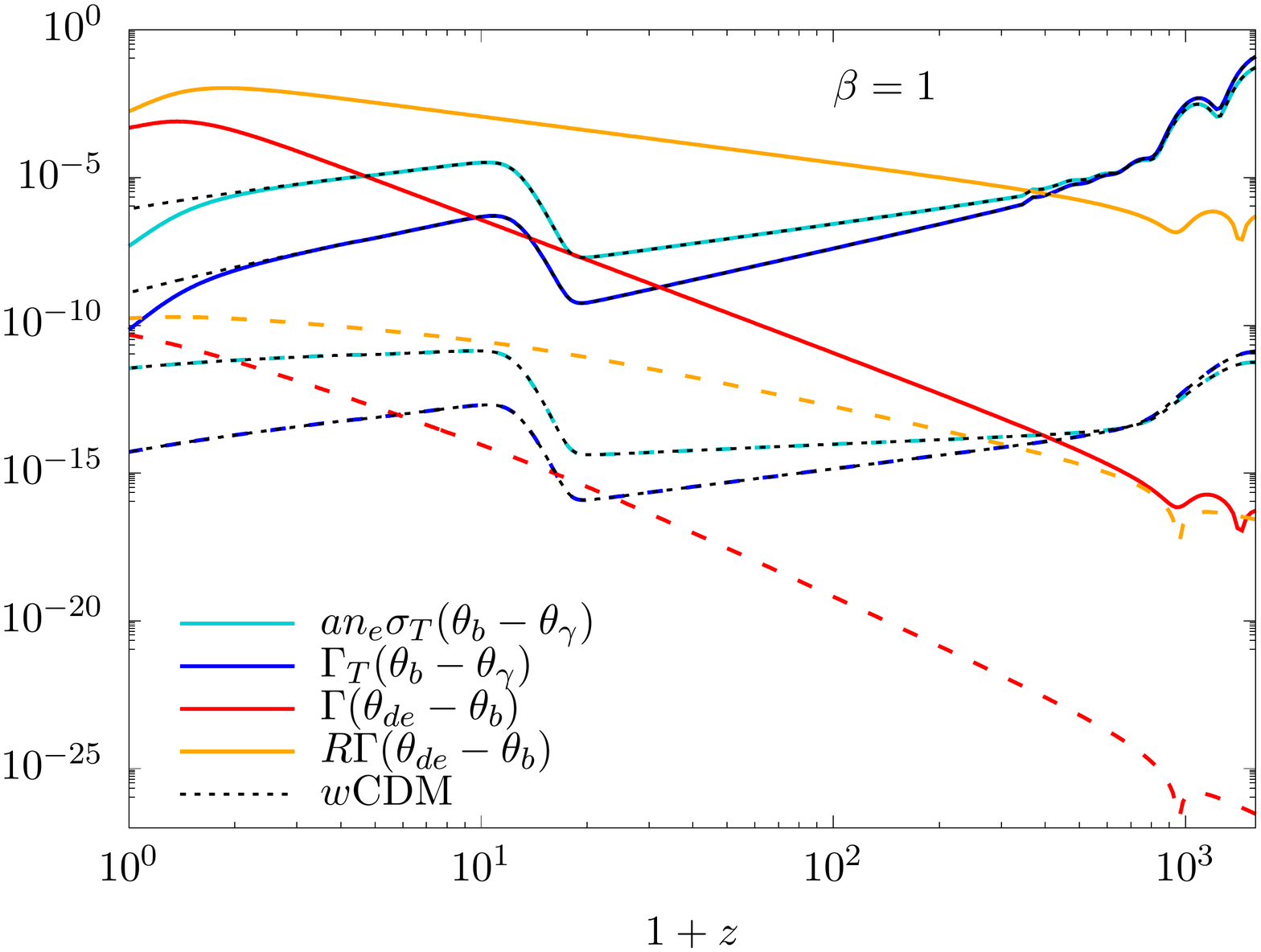}
		\includegraphics[width=0.5\textwidth]{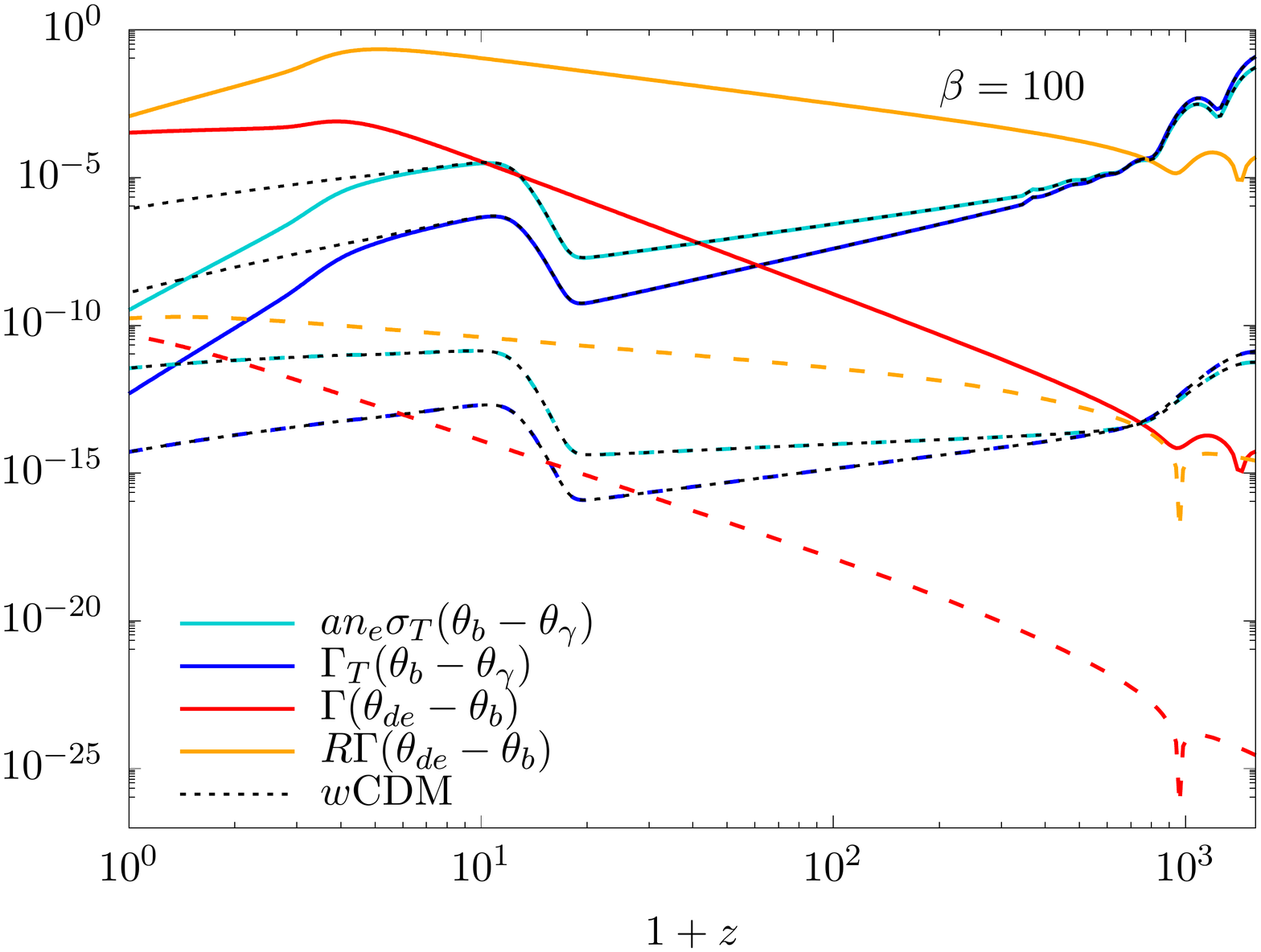}}
	\caption{Comparison of the interaction terms of DE-baryons and baryons-photons coupling fluids throughout the cosmological evolution, for modes $k=10^{-1}\,$Mpc$^{-1}$ and $k=10^{-4}\,$Mpc$^{-1}$ (solid and dashed lines, respectively), for $\beta=1,100$. Dashed black lines represent the corresponding $w$CDM case ($\beta=0$), included for comparison. In the right panel with higher value of $\beta$ we can see that the interaction of baryons with DE can be more important than the Thomson scattering with potential effects for the reionisation epoch. For most of our values of $\beta$ and explored scales, this is not relevant however and our approximation in Section \ref{sec:analytical} is justified.}
	\label{fig:evolTERMSz}	
\end{figure}

 The evolution in redshift of the interaction terms involved in the equations for velocities of baryons, photons and DE is shown in Fig.~\ref{fig:evolTERMSz} for $\beta=1,100$. We choose  the representative Fourier modes $k=10^{-1}\,$Mpc$^{-1}$ and $k=10^{-4}\,$Mpc$^{-1}$ indicated with solid and dashed lines respectively and also include the $w$CDM case ($\beta=0$) for comparison. As already mentioned, the coupling terms become relevant at low redshift for both values of $\beta$. While large scales are not affected since the fluids share a common rest frame, for higher values of $k$ the strength of Thomson scattering terms is substantially dropped due to the deviation of the velocity of baryons from the standard behaviour (see Fig.~\ref{Fig:velcoupled} below).

\subsection{Observables: Matter power spectrum and CMB}
\label{sec:pkCMB}
\begin{figure}[!t]
	\centerline{
		\includegraphics[scale=0.120]{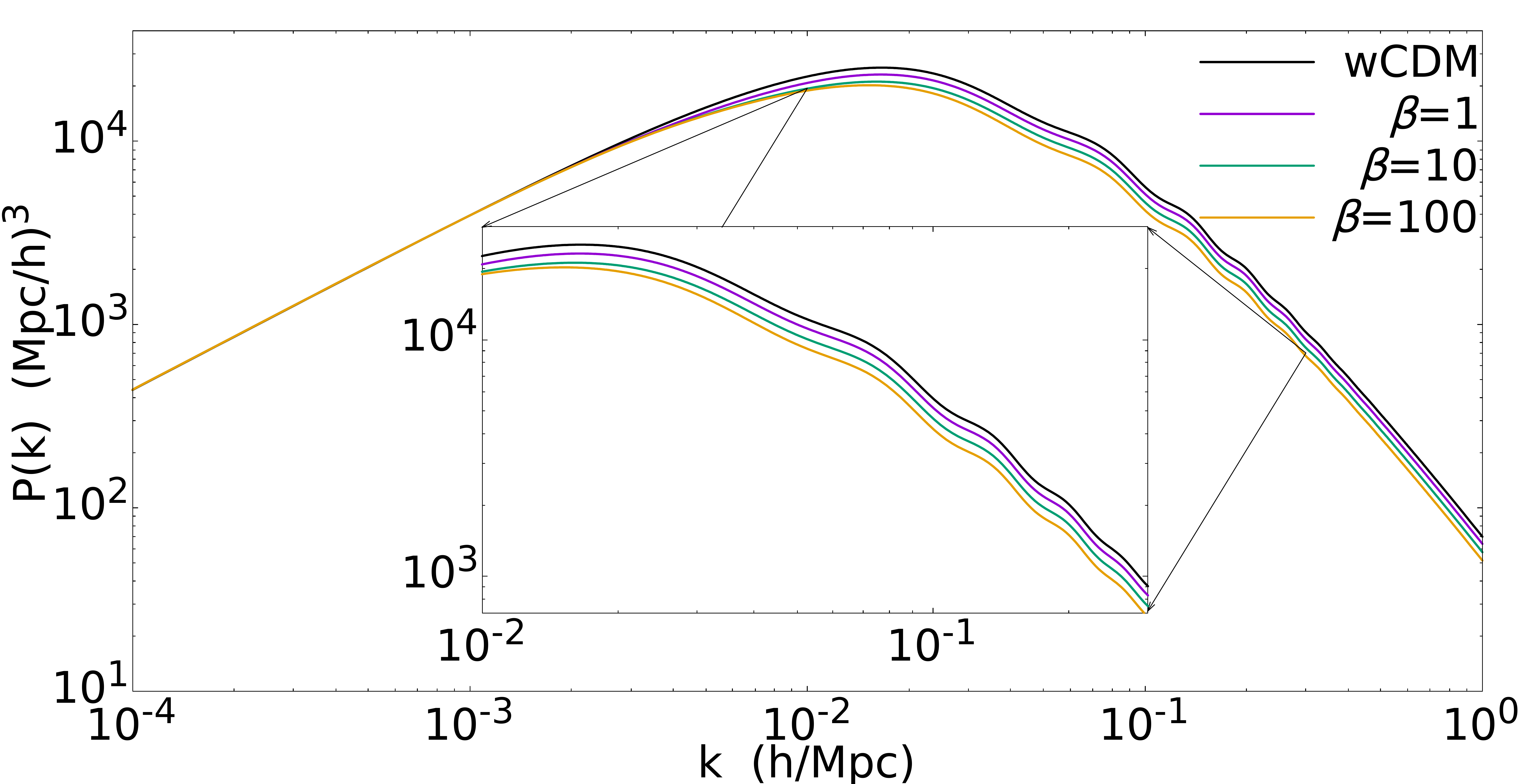}
	    \includegraphics[scale=0.120]{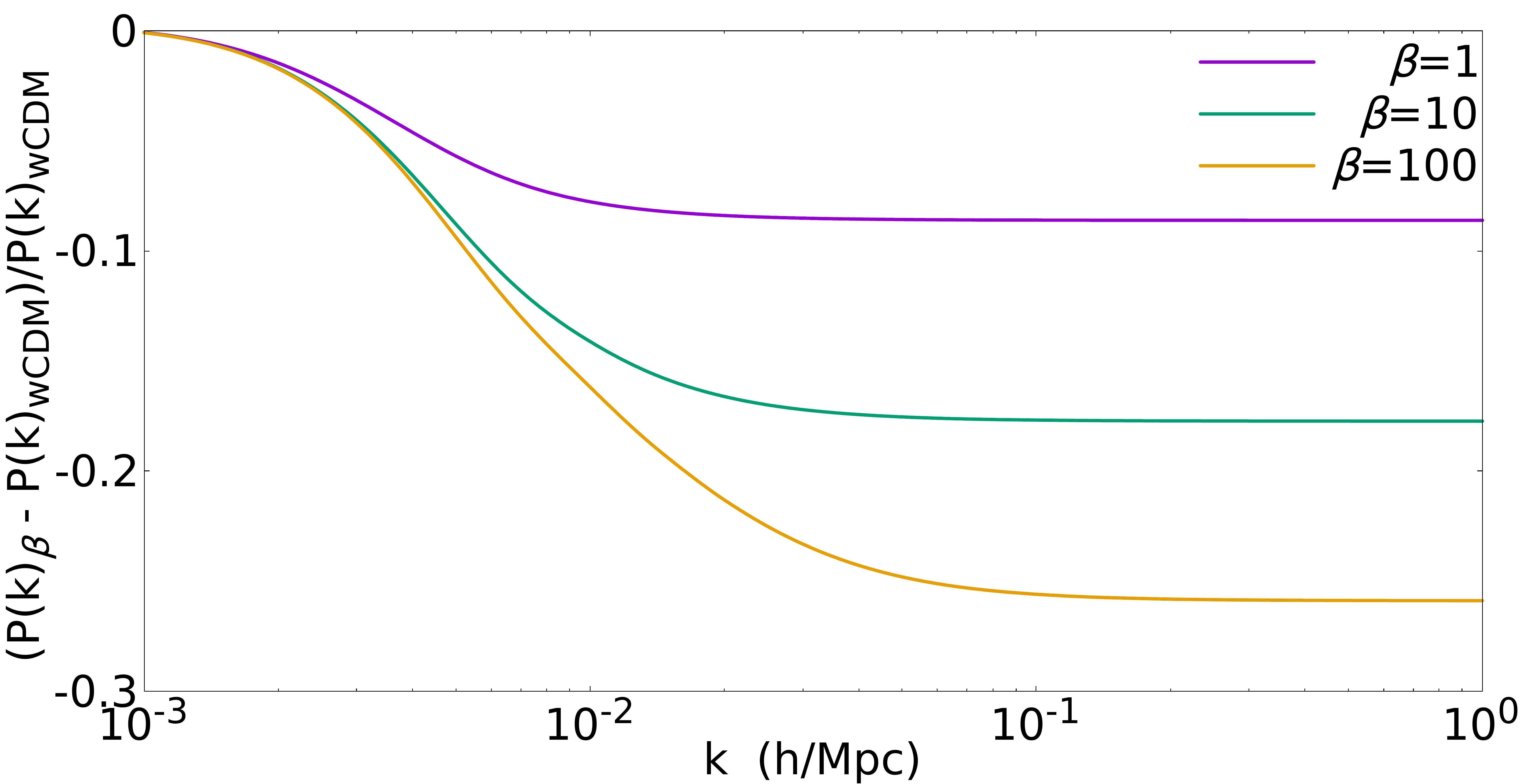}}    
	\caption{In the left plot, we show the matter power spectrum for the reference model $w$CDM and for several values of the coupling parameter $\beta$ for the interacting model. In the right plot, we show the relative ratio for several values of $\beta$.}
	\label{Fig:Pk}	
\end{figure}

As we explain in Section~\ref{sec:analytical}, the baryons-DE elastic interaction causes both components to form a locked system, forcing the growth of baryonic structures to deviate from the standard one, when the interaction is efficient. Specifically, the density contrast becomes constant as we will see in Section~\ref{sec:dbs8} and in accordance with our analytical result. In Figure~\ref{Fig:Pk}, we show how this lock leaves a precise imprint in the matter power spectrum having different regimes depending on the scale. As one would expect, on large scales the interaction has no influence since both DE and baryons have the same rest frame, thus the coupling term vanishes. When we consider smaller scales we have two different regimes, on intermediate scales we see the previously explained effect as a $k$-dependent suppression on the matter power spectrum, exhibiting more suppression as we go to smaller scales. For small enough scales, this suppression saturates and becomes $k$-independent since the gravitational pull dominates over the DE drag. In all scenarios, the suppression increases with $\beta$ since this parameter measures the strength of the interaction. 

We can infer an important feature from the behaviour of the matter power spectrum, the peak is no longer only determined by the scale of matter-radiation equality, but it suffers a shift because of the interaction. This effect is due to the lock of DE and baryons leading to a suppression of the growth of structures, that is only relevant for small enough scales. Even more remarkable is that such shift is obtained without modifying the background cosmology or the matter content of the Universe, therefore, it constitutes a very distinctive feature of the elastic interaction. This effect also appears in Ref.~\cite{Asghari:2019qld}, where a similar interacting term is used but, in that case, coupling DE to DM instead of baryons.

\begin{figure}[!t]
	\centerline{\includegraphics[scale=0.3]{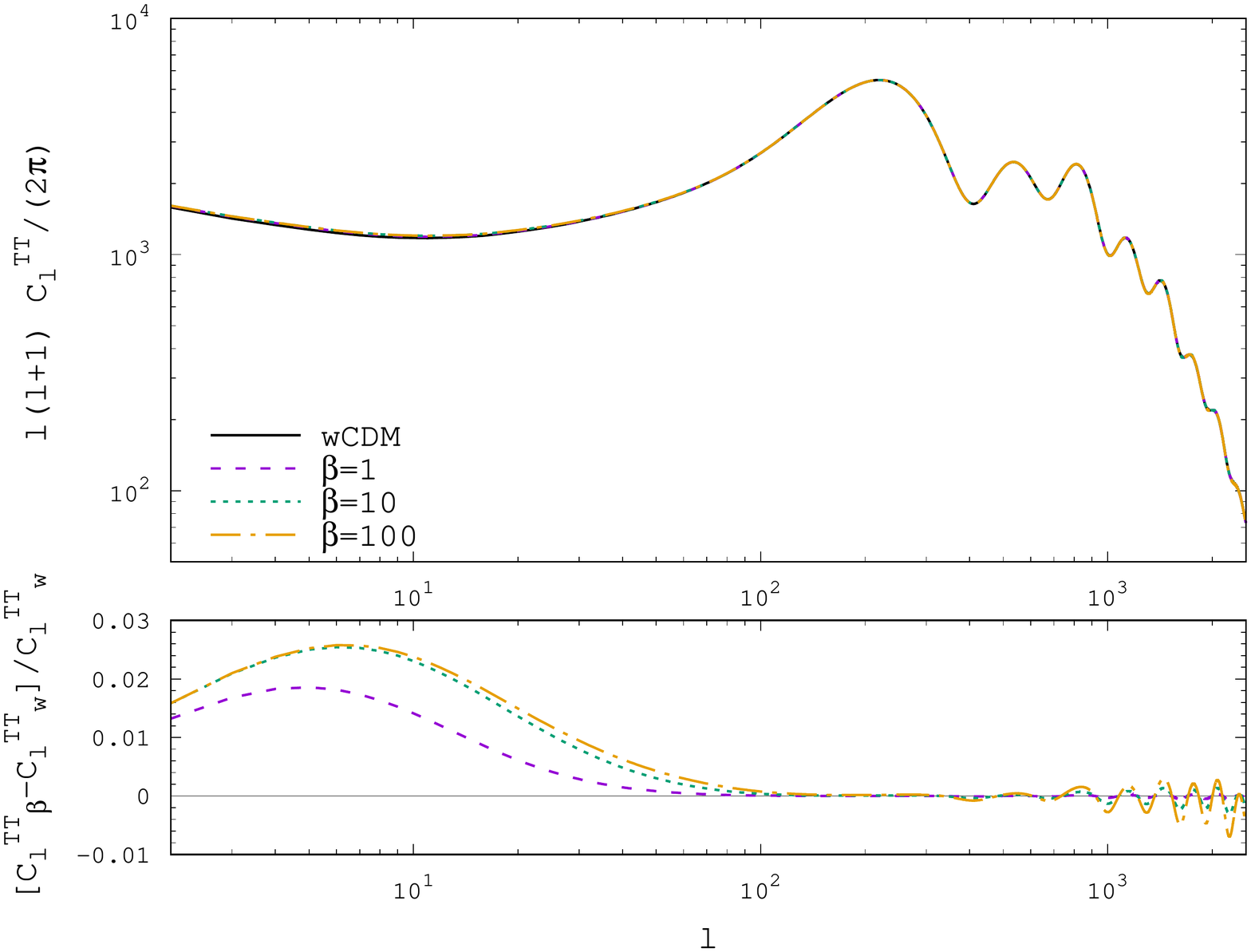}
	\includegraphics[scale=0.3]{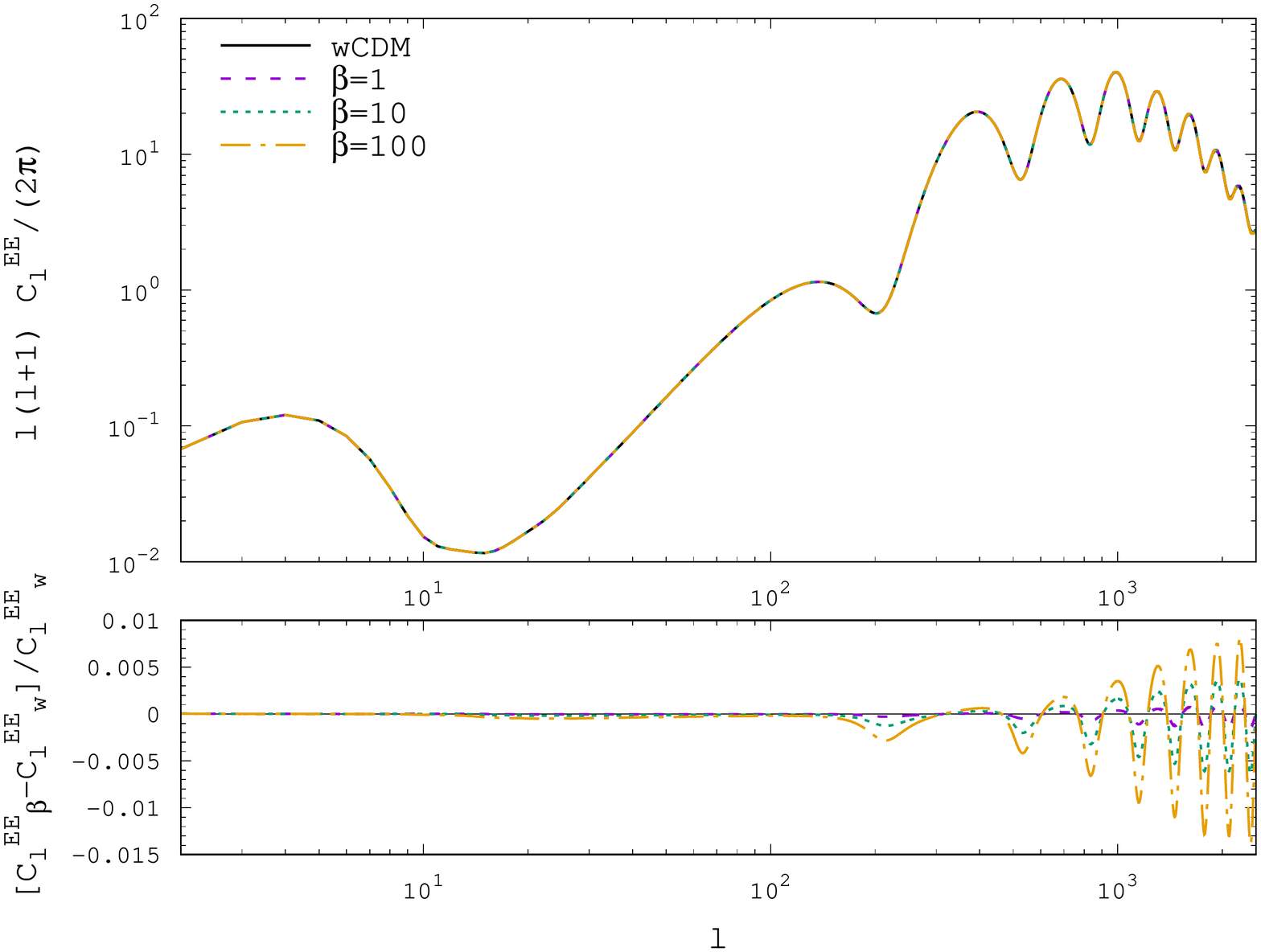}}
	\centerline{\includegraphics[scale=0.3]{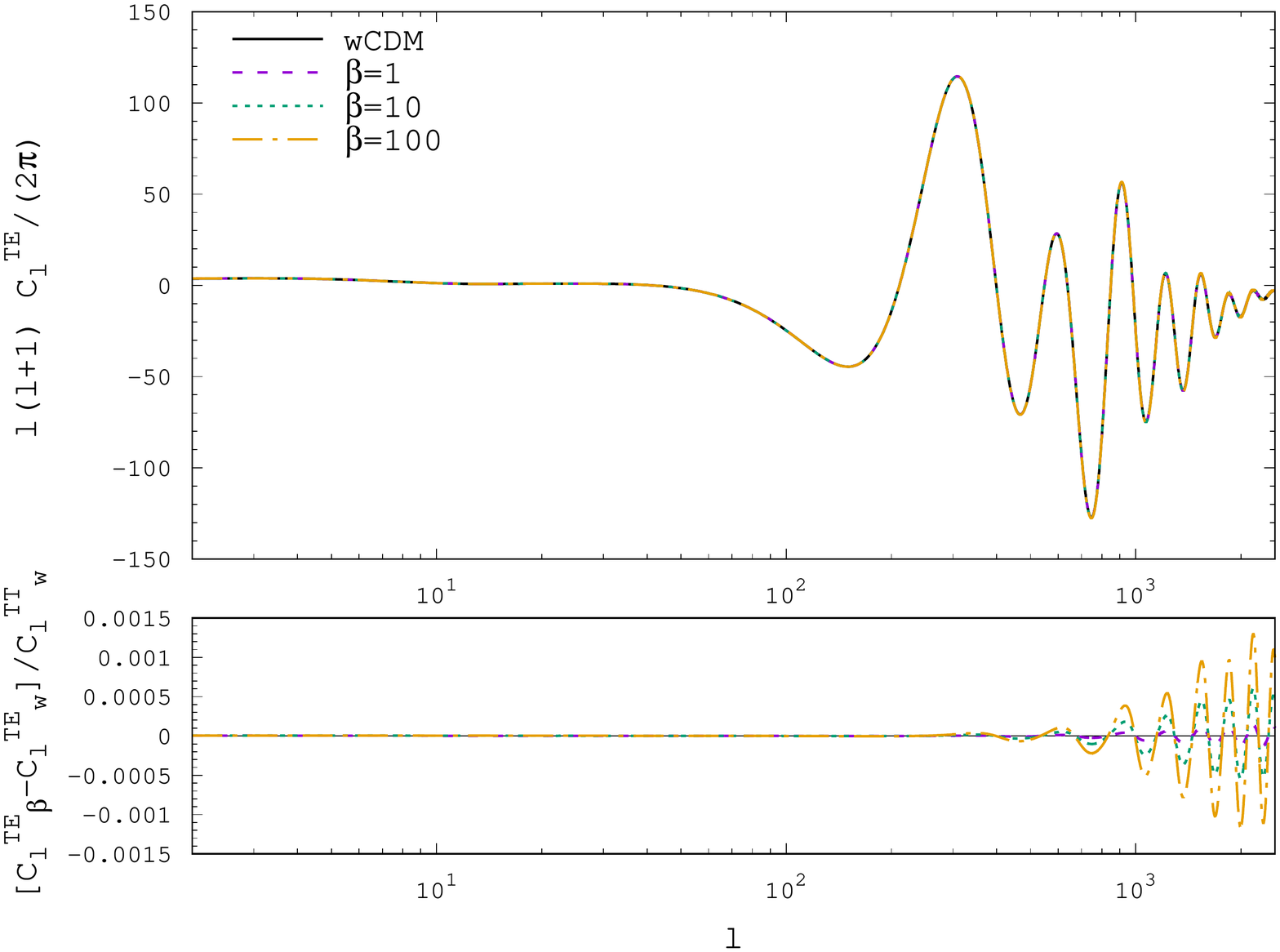}
	  \includegraphics[scale=0.3]{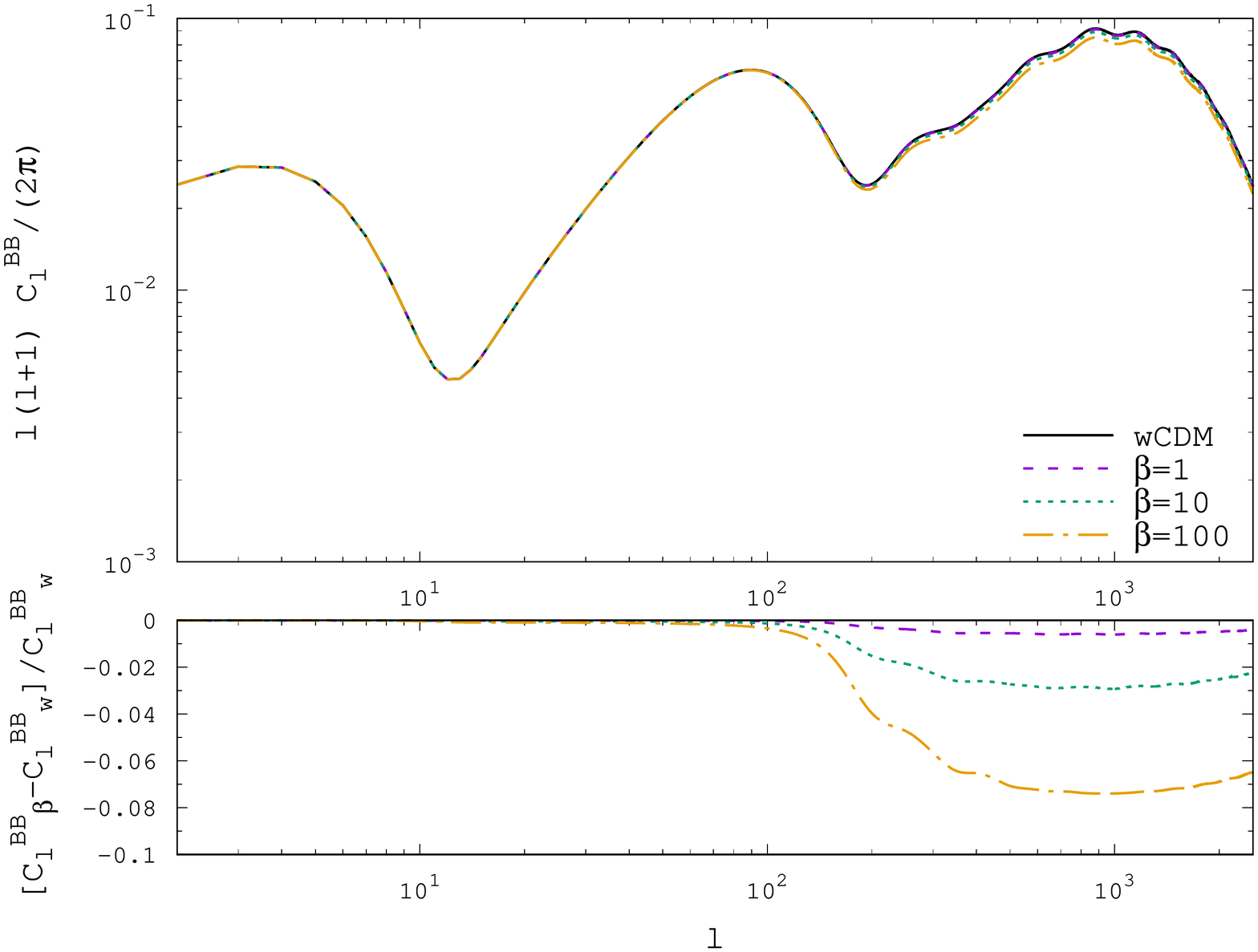}} 
	\caption{CMB angular power spectra for different values of the model parameter $\beta$. In all cases, the normalisation of relative errors is given w.r.t. $w$CDM model ($\beta=0$).}
	\label{fig:Cls}	
\end{figure}

\begin{figure}[!t]
	  \centerline{\includegraphics[scale=0.3]{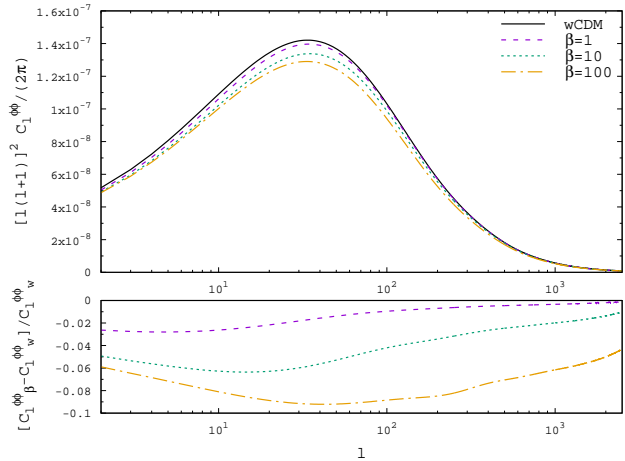}}
	\caption{Lensing potential for different values of the parameter $\beta$. The amplitude of the lensing potential diminishes for higher values of $\beta$, leading to higher $l$ oscillations in the CMB temperature and polarisation power spectra.}
	\label{fig:phiCls}	
\end{figure}

The CMB power spectra for temperature, polarisation and cross-correlations for the scenario we are considering are shown in Fig.~\ref{fig:Cls}  for different values of the interaction parameter $\beta$ together with those for the $w$CDM model. 
The temperature angular power spectrum is mainly modified on large scales, as expected because the interaction is  relevant at very late times via late-time Integrated Sachs-Wolfe effect. However, the interaction also affects the power spectra at small scales through the effect of lensing, which leads to high-$l$ oscillations present in both temperature and polarisation, with the exception of the $BB$ power spectrum that has a non-oscillating correction at high $l$. 
As shown in Fig.~\ref{fig:phiCls}, the parameter $\beta$ diminishes the amplitude of the lensing potential $C_l^{\phi\phi}$ and the higher $l$ oscillations.

\subsection{Suppression of structures}
\label{sec:dbs8}

As we have discussed in the previous sections,  there is a regime in which the interaction term is such that baryons and DE form  a locked system. This coupled system causes DE to drag baryons, preventing them to fall into the DM potential wells, and hence reducing its clustering amplitude. The consequence of this effect appears already in the matter power spectrum as a suppression on intermediate and small scales, as can be seen in Figure \ref{Fig:Pk}. We will now discuss this point in a more explicit way by studying the evolution of the density perturbation and by exploring the consequences on the parameter $\sigma_8$.

\begin{figure}[!t]
	\centerline{
		\includegraphics[scale=0.120]{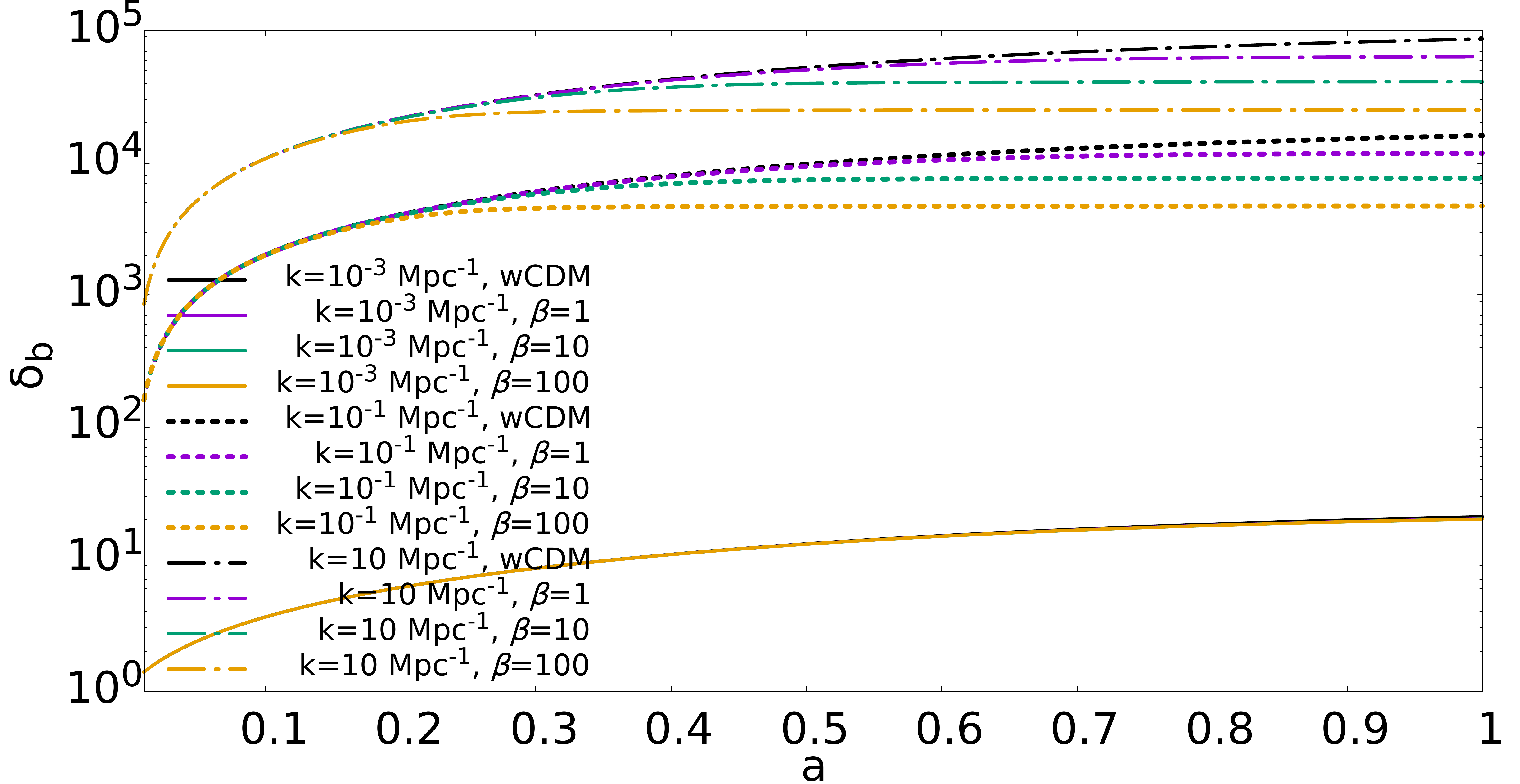}
	    \includegraphics[scale=0.120]{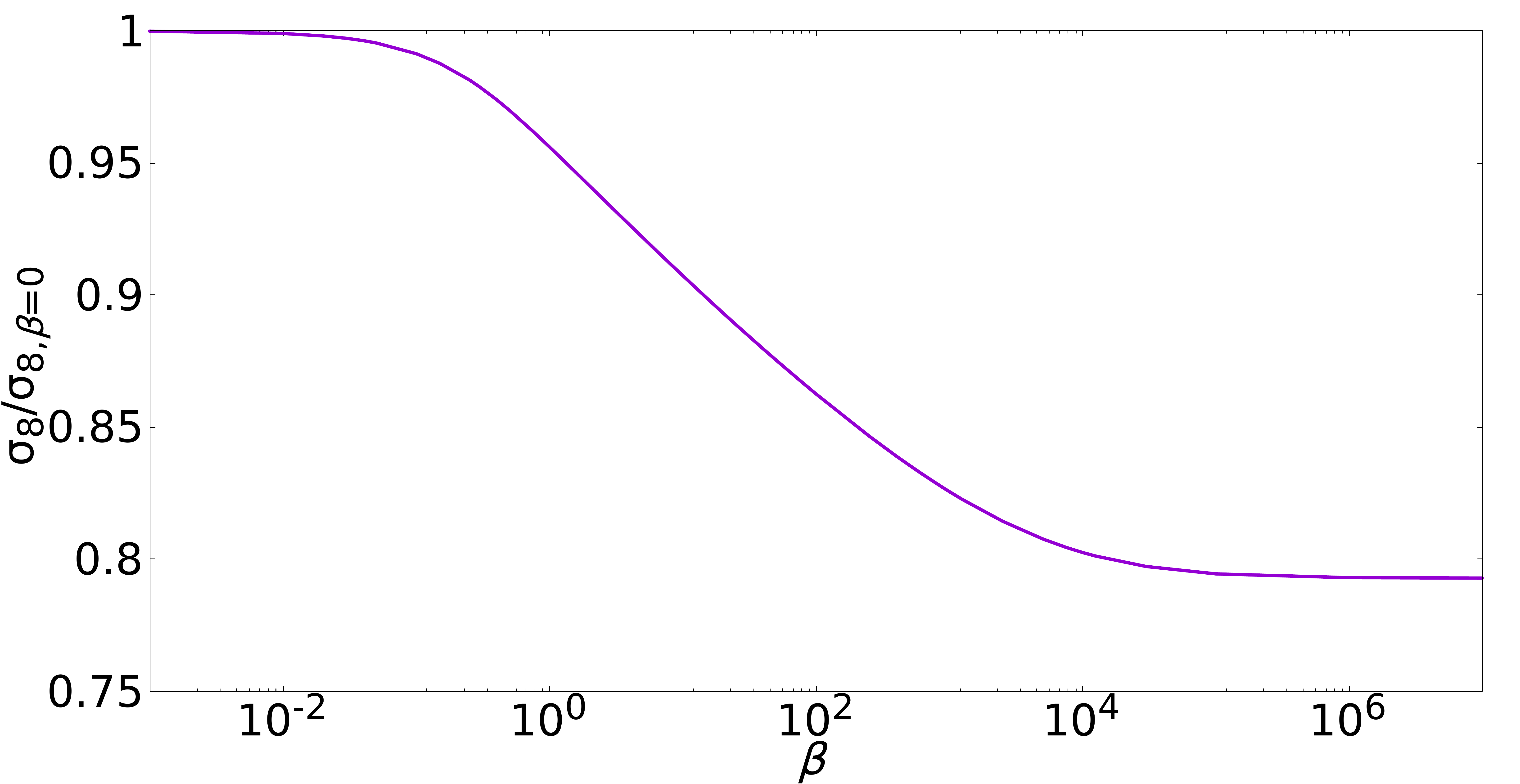}}    
	\caption{In the left plot, we show the evolution of the baryon density contrast $\deltab$ for several modes and several values of the coupling parameter $\beta$. The solid lines represent the mode $k=10^{-3}$\,Mpc$^{-1}$, non perturbed by this kind of interaction, the dashed lines the mode $k=10^{-1}$\,Mpc$^{-1}$ and the dash-dotted lines the mode $k=10$\,Mpc$^{-1}$, both affected by the interaction. The black lines represent the reference model $w$CDM while the purple, green and yellow lines are the interacting model with $\beta=1$, $10$ and $100$, respectively. 
	In the right plot, we explicit the value of the $\sigma_8$ parameter, with respect to its non interacting value $\sigma_{8,\beta=0}$ , depending on $\beta$ for the same cosmological parameters.}
	\label{Fig:structures}	
\end{figure}

As baryons are dragged by DE, we can understand such suppression by looking at the evolution of the baryon density contrast $\deltab$ when the elastic interaction is efficient, that is, at late times and sub-horizon scales. In the left plot of Figure~\ref{Fig:structures}, we see, as anticipated, that the density contrast for the large scale mode $k=10^{-3}$\,Mpc$^{-1}$ (solid lines) experiences no  suppression since both components have the same rest frame and, therefore, the interacting term is inefficient. The intermediate (dotted lines) and small scale modes (dash-dotted lines) experience a suppression of the growth of structures as DE starts to drag baryons from $z\sim 2$ on. When the interaction becomes more efficient, the baryonic growth of structures freezes, thus deviating from the standard growth of the late Universe. We find this suppression is not only  more significant as the value of the coupling parameter $\beta$ increases, but it also starts earlier, following also the tendency inferred from Section~\ref{sec:regimes}. Although the freezing of the baryon density contrast $\deltab$ can lead to a variation in the Newtonian potential $\Phi$, for the values of the coupling parameter $\beta$ considered here and given that the dominant contribution to  $\Phi$ comes from DM, such variation is negligible. Hence, we can take $\Phi\sim const$ well inside matter domination even after the interaction turns on, thus supporting our assumptions in Section~\ref{sec:analytical}.

As a consequence of the freezing of the growth of baryonic structures, there is less matter clustered if we increase the value of the coupling constant $\beta$. This leads to an imprint on the $\sigma_8$ parameter, getting a lower value as we show in the right plot of Figure~\ref{Fig:structures}. For a large  range of values of the coupling parameter $\beta\in [0.1,10^4]$ we have different values of $\sigma_8$, but as the baryonic matter dilutes and the dragging reaches its maximum efficiency, its value saturates at the value $\sigma_8\sim \frac{4}{5}\,\sigma_{8,\beta=0}$, with $\sigma_{8,\beta=0}$ the non interacting value, which is mainly determined by DM. This saturation can be understood in terms of the small fraction of baryons that contribute to the total matter clustering. We recall that such suppression is achieved without changing the background cosmology or the energy/matter content of the Universe and, therefore, it can alleviate the current $\sigma_8$ tension. This allows to improve the fit to observational data that we will perform below.

\subsection{Induced relative velocities between matter components and dark energy - baryons oscillations}

In the standard cosmological scenario, the velocities of DM and baryons, after they are decoupled from photons, evolve similarly since both components are falling into the potential wells. In the present scenario, however, due to the fact that the interaction is mainly determined by the relative velocity between DE and baryons this is no longer the case. We remind that the relative velocity is gauge independent and for the Figures that show individual velocities we use the Newtonian gauge, where matter velocities dominate over the DE one (suppressed due to its high pressure), with the exception of large scale where all have the same rest frame.

\begin{figure}[!t]
	\centerline{
		\includegraphics[scale=0.120]{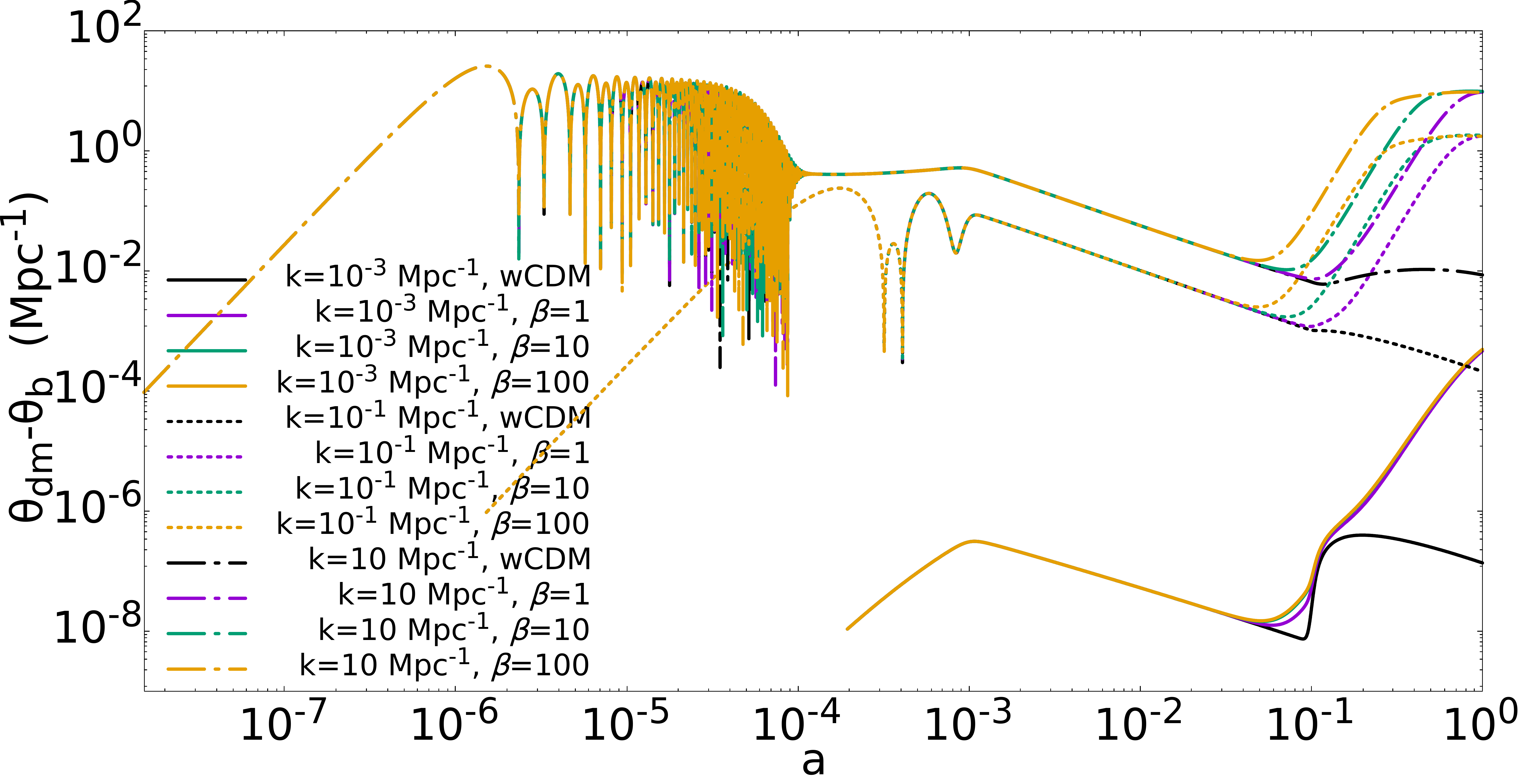}
		\includegraphics[scale=0.120]{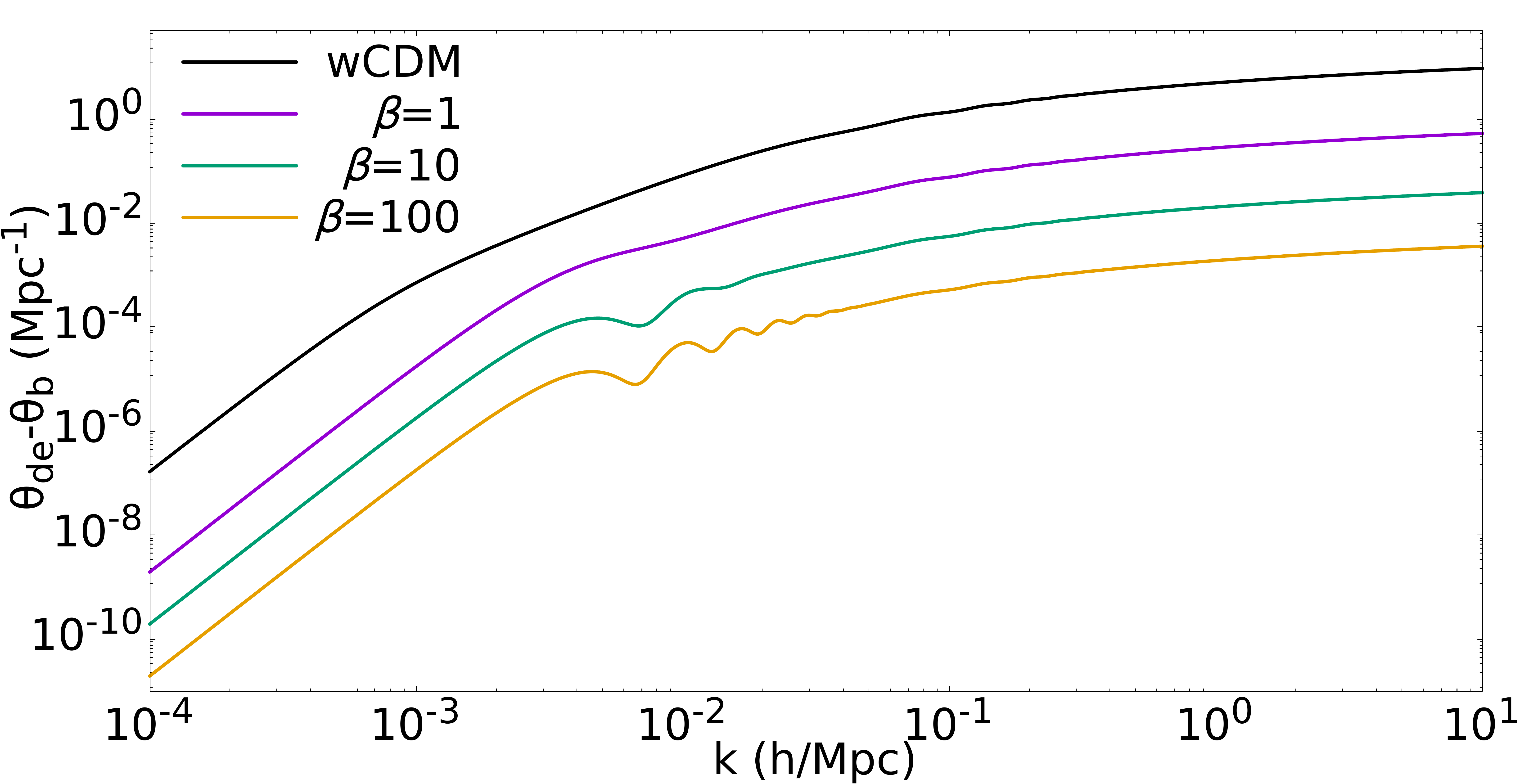}}
	\caption{In the left plot, we show the evolution of the the relative velocity between DM and baryons. In the right plot, we display the relative velocity today between DE and baryons for different scales. The black line represents the reference model $w$CDM while the purple, green and yellow lines are the interacting model for the values of the  coupling parameter $\beta=1$, $10$ and $100$, respectively. As before, the solid lines represent the mode $k=10^{-3}$\,Mpc$^{-1}$, the dashed lines the mode $k=10^{-1}$\,Mpc$^{-1}$ and the dash-dotted lines the mode $k=10$\,Mpc$^{-1}$.}
	\label{Fig:DAO}	
\end{figure}

\begin{figure}[!t]
	\centerline{
		\includegraphics[scale=0.120]{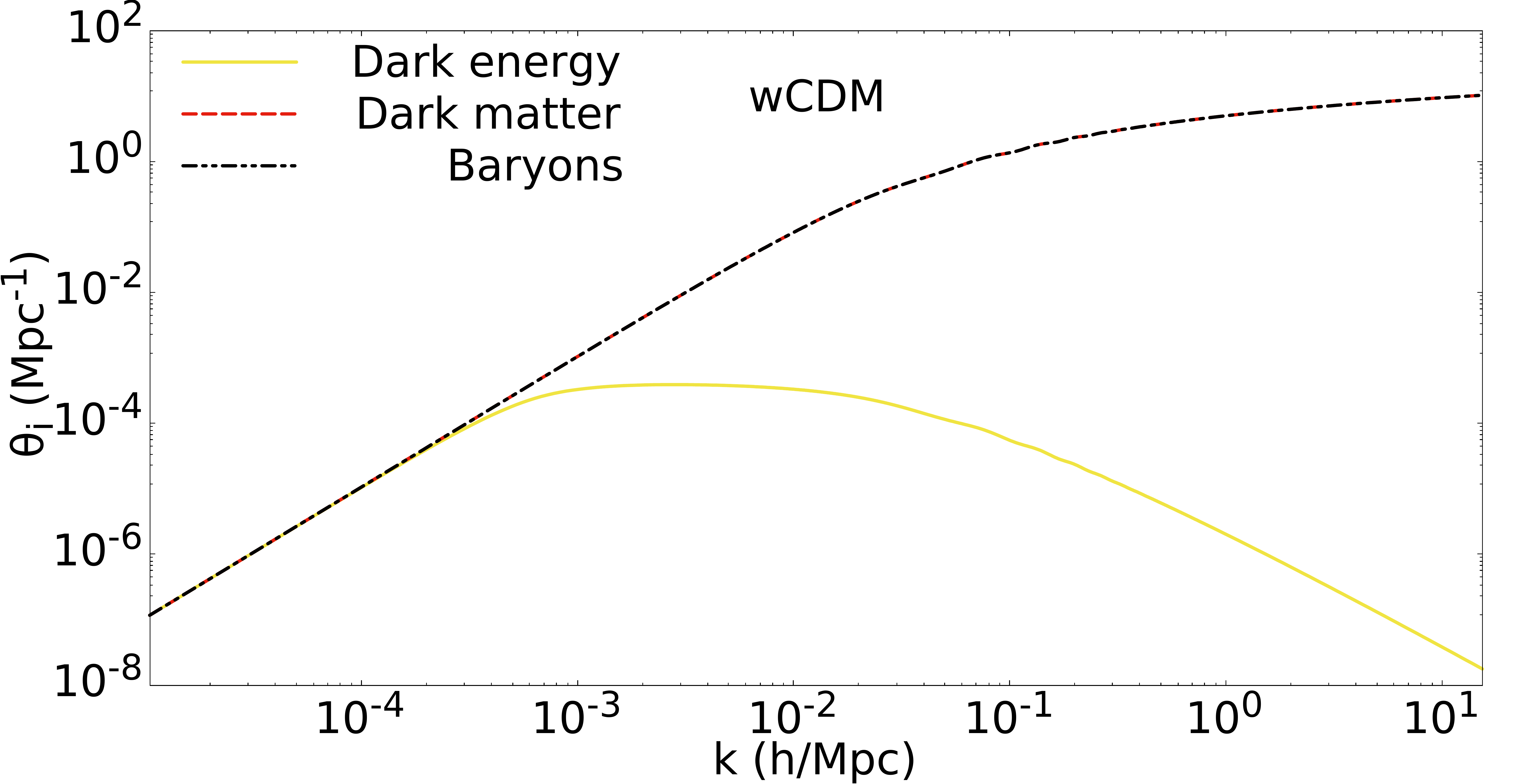}
		\includegraphics[scale=0.120]{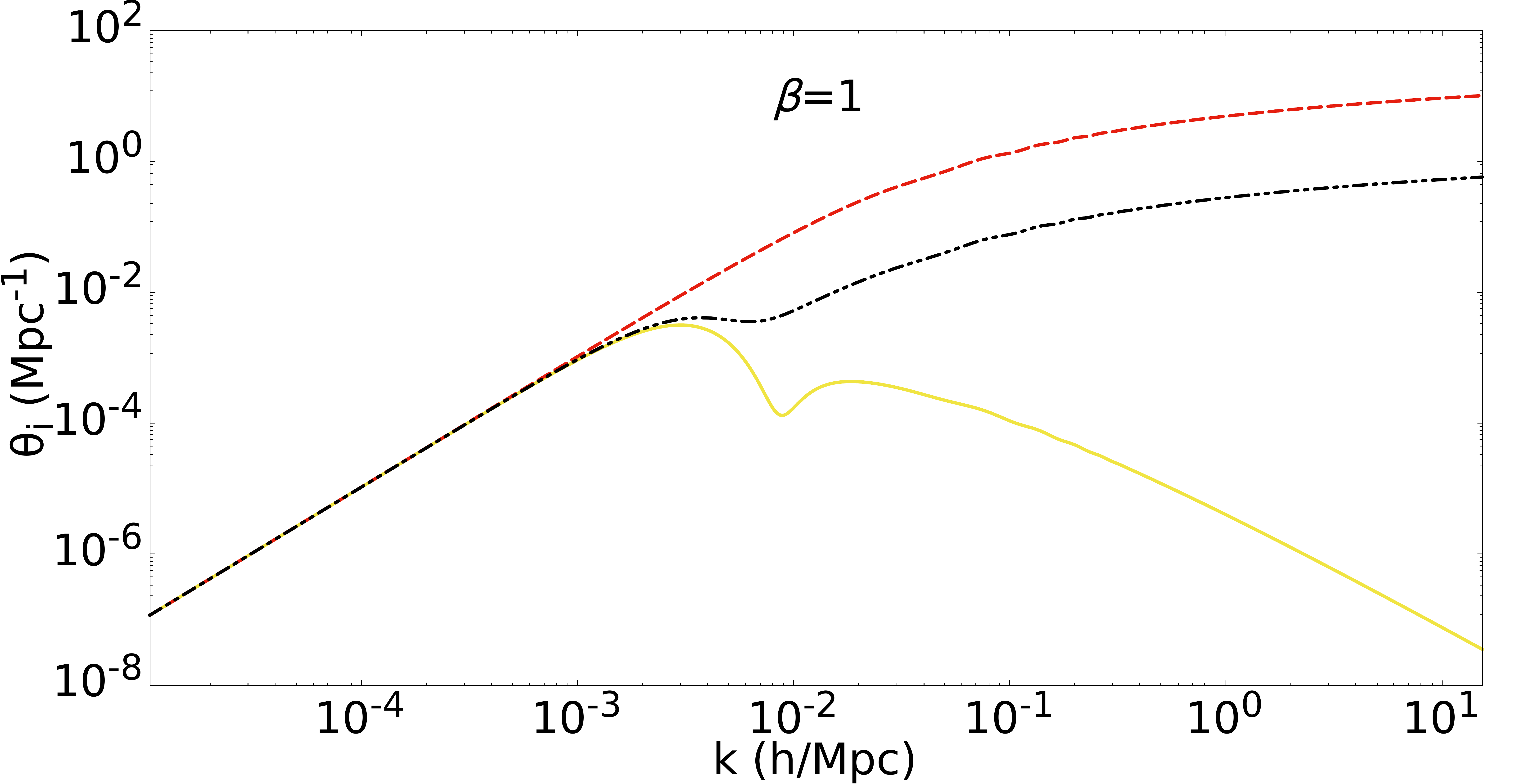}}
		\centerline{
		\includegraphics[scale=0.120]{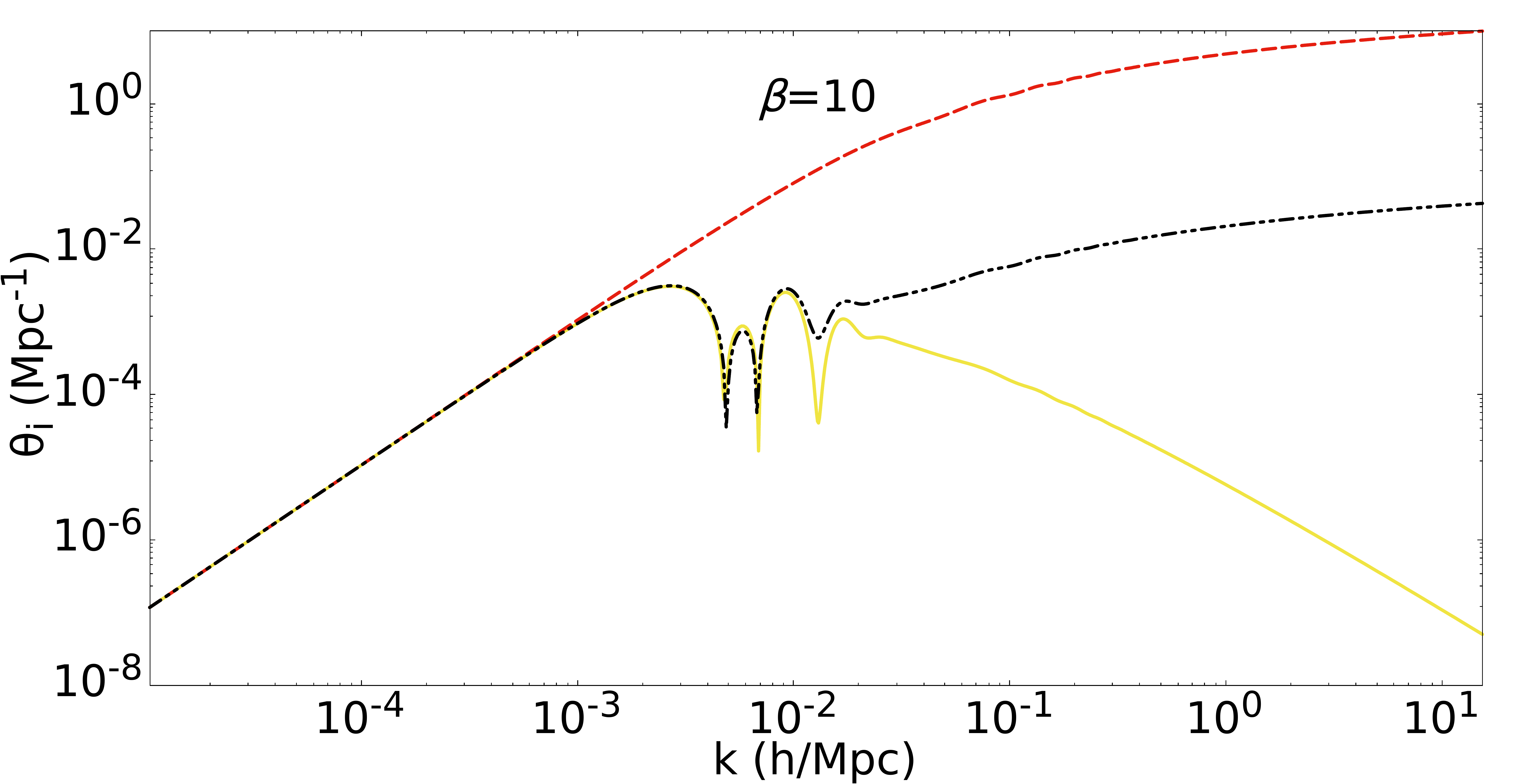}
		\includegraphics[scale=0.120]{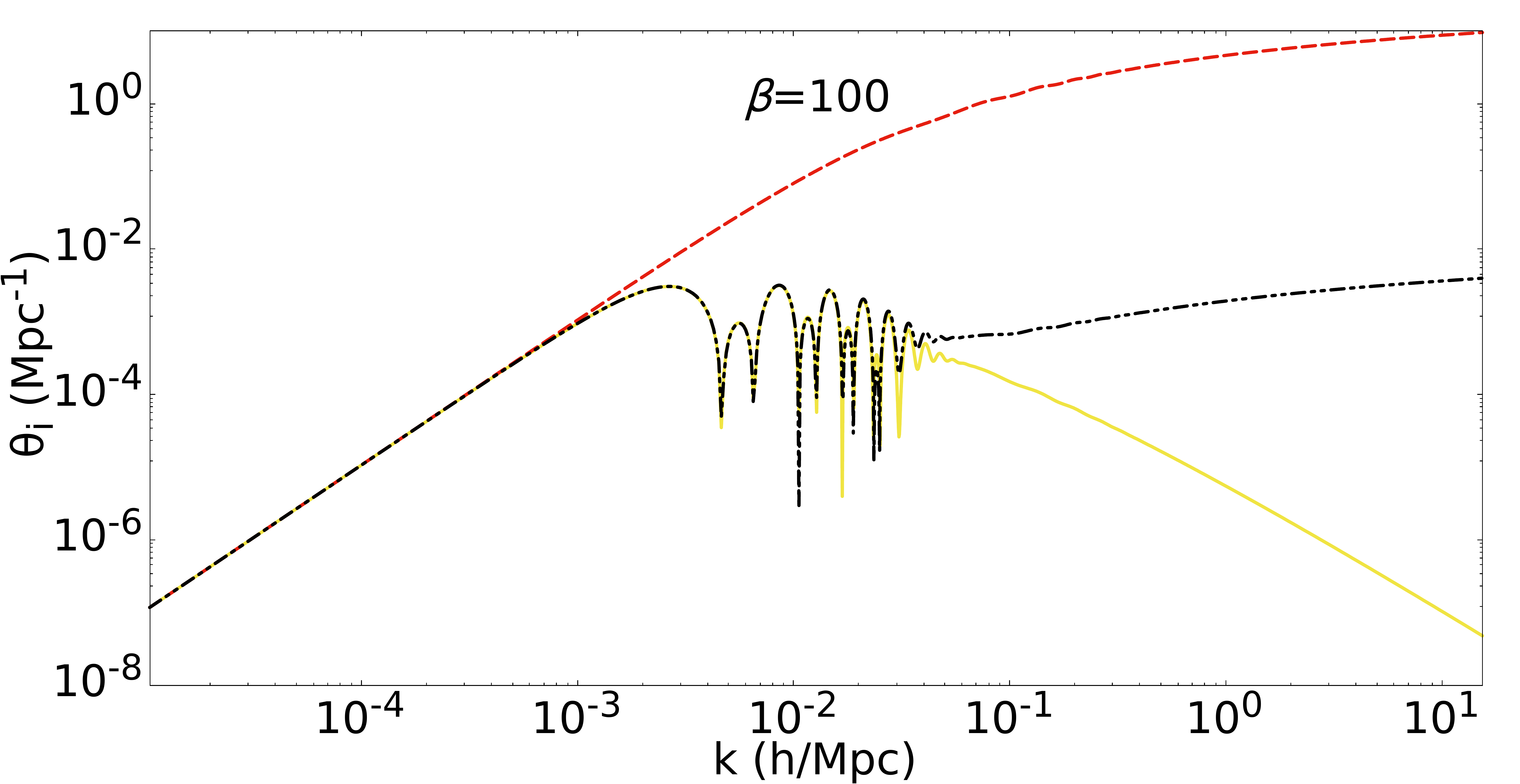}}
	\caption{In these plots, it is shown the $k$ dependence of today's velocity of DE (yellow line) and baryons (black line) for various values of the coupling constant $\beta$.  As a reference for baryons, we also plot the unaffected DM velocity (red line). We see that large scale behaviour is unchanged for every value of the  coupling parameter, since the interacting term vanishes. On intermediate scales the oscillations appear, whose number depends on the value of $\beta$. This shows the competition between the DE drag and the gravitational pull, resulting in a suppression of baryons' velocity. Finally, at small scales, baryons and DE velocities decouples as consequence of the gravitational pull that dominates over the DE drag. These plots are done using the Newtonian gauge, where matter velocities dominate over DE ones.}
	\label{Fig:velcoupled}	
\end{figure}

In the left plot of Figure~\ref{Fig:DAO}, we display the evolution of the relative velocity between DM and baryons and how they get a new induced one when the interaction becomes efficient at late times. While in the right plot, we show how the relative velocity today between DE and baryons is lower when the interaction is switched on, as it tends to couple both. These effects are also inferred from  Figure~\ref{Fig:velcoupled}, where we show how, when the interaction is on, the velocity of baryons deviates today from the DM one at intermediate and small scales, trying to get coupled to the DE velocity. This new relative velocity is the responsible for the suppression of structures shown before in the matter power spectrum or in the baryon density contrast, since it causes that baryons do not fall into the gravitational wells following DM, but they are partially dragged by DE.

One characteristic feature on velocities of this  model is the intermediate scale regime. As previously described in Section~\ref{sec:analytical}, it exhibits the coupled oscillations in the DE and baryons velocity. We see on that scales the competition between the gravitational pull, trying to make baryons fall into the potential wells created by DM, and the drag of the elastic interaction inducing baryons to couple DE. On small enough scales, the drag induced by DE on baryons cannot compete with the gravitational collapse and it is only capable of slowing it down.

\section{Observational constraints}
\label{sec:fit}
Now that we have explored the possible effects of the elastic interacting model, we turn our attention to its compatibility with observations. We use the public code of Markov chains Monte Carlo called MontePython \cite{Brinckmann:MP,Audren:MP} applied to our modified CLASS code, to fit several cosmological parameters and our coupling parameter with available data. We use the full Planck 2018 dataset~\cite{PLANCK2018,Planck2018LK} containing data of high-l  and low-l from CMB temperature (TT), polarisation (EE), the cross correlation of temperature and polarisation (TE) and the CMB lensing power spectrum, the JLA likelihood with supernovae data~\cite{JLA}, the BAO combined data~\cite{BAO1,BAO2,BAO3}, the likelihood of data measured with Planck of the Sunyaev-Zeldovich effect~\cite{PlanckSZ} and the likelihood from weak lensing data CFHTLenS~\cite{CFHTL}. 

For our analysis, we consider as cosmological parameters the baryon density defined as $100~\Omega_{\rm b} h^2$, the DM density $\Omega_{\rm dm}h^2$, the scalar spectral index $n_{s}$, the primordial amplitude $10^{9}A_{s}$, the reionisation optical depth $\tau_{\rm reio}$, the equation of state of DE $w$ (constrained to the non-phantom region $w>-1$) and the angular acoustic scale as $100~\theta_{s}$. In addition, we consider as derived parameters the redshift of reionisation $z_{\rm reio}$, the Hubble parameter $H_0$, the  matter fluctuation amplitude at $8h^{-1}$Mpc\, as $\sigma_8$ and the total matter density $\Omega_{\rm m}$. We also fix the DE sound speed to $\cs^2=1$.

Finally, the cosmological parameter associated to the interaction is $\beta$. We have normalised this parameter so its natural value is ${\mathcal{O}}(1)$. Although this might be a natural guess from a theoretical point of view, we do not have any external information on this parameter. For that reason, we will consider two classes of priors, namely: a flat prior on ${\log_{{10}}\,  \beta }$ and a flat prior on $\beta$. We will see that the former selects the expected natural order of magnitude which is confirmed by the latter, although with a poorer convergence.

\subsection{Flat prior on $\log_{10}\, \beta$}

We start our analysis by considering a flat prior on ${\log_{{10}}\,  \beta }$ over the range ${\log_{{10}}\,  \beta }\in[-8,4]$. A potential caveat of this prior is that we may be artificially excluding the non-interacting case $\beta=0$. However, as we will see the $2\sigma$ contour is entirely contained within this region so we can be confident that the non-interacting case is not excluded by a flawed choice of prior. Furthermore, notice that values of $\beta$ smaller than $10^{-8}$ do not give any appreciable deviation with respect to the non-interacting case however so our range is safe.

\begin{figure}[!t]
	\centerline{
		\includegraphics[scale=0.47]{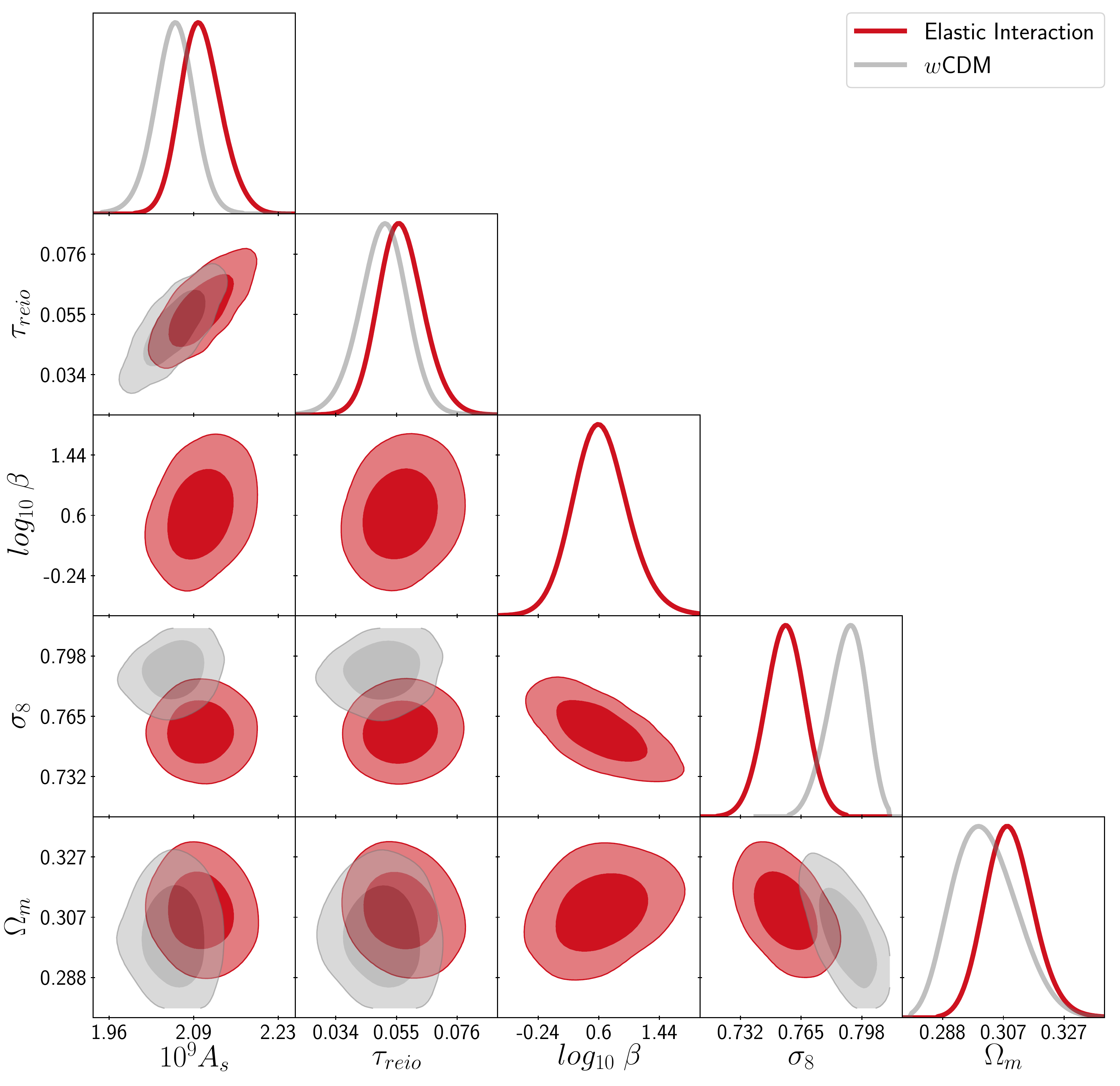}}
	\caption{The one-dimensional posterior  distributions and the two-dimensional contours obtained for several parameters for the $w$CDM model (gray) and the interacting model (red), using the prior ${\log_{{10}}\,  \beta } \in [-8,4]$. We can clearly see the shift in $\sigma_8$ caused by the interaction and that permits to improve the $\chi^2$ value.}
 	\label{fig:fit}	
\end{figure}

In Table~\ref{tab:fit}, we show the mean with $1\sigma$ confidence limits and the $2\sigma$ upper and lower limits for the cosmological parameters, the derived parameters and the coupling parameter, for a $w$CDM model and the interacting model obtained with the previously explained datasets. In Figure~\ref{fig:fit}, we display the one-dimensional  posterior  distributions  and  the  two-dimensional  contours obtained  for  several  parameters. We find a slight decrease in the baryon density $100~\Omega_{\rm b} h^2$ and an increase in the DM density $\Omega_{\rm dm}h^2$, that translates into a slightly higher value of the total matter density $\Omega_{\rm m}$, if we compare with a $w$CDM model. A significant difference appears in the $\sigma_8$ parameter that has a smaller value, consistently with  the right plot of  Figure~\ref{Fig:structures}, and it is closely connected to the suppression of structures that the interaction provokes. It is worth stressing that all these differences are achieved with no modification of the background cosmology. Finally, we find a $\Delta\chi^2=26$ improvement with respect to the $w$CDM model.

\begin{table}[!t]
\begin{center}
\renewcommand{\arraystretch}{1.8}
\begin{tabular}{ |c||c|c|c||c|c|c| } 
	\hline
	\hline
	\centering
	&\multicolumn{3}{c||}{$w$CDM model} &\multicolumn{3}{c|}{Elastic Interaction}\\ \hline
	Param. & mean$\pm\sigma$ & $2\sigma$ lower & $2\sigma$ upper  & mean$\pm\sigma$ & $2\sigma$ lower & $2\sigma$ upper\\
	\hline \hline \hline
$100\Omega_{\rm b } h^2$ & $2.264_{-0.015}^{+0.015}$ & $2.235$ & $2.294$ & $2.243_{-0.016}^{+0.016}$ & $2.211$ & $2.275$ \\\hline
$\Omega_{\rm dm}h^2$  & $0.1163_{-0.001}^{+0.001}$ & $0.1143$ & $0.1183$ & $0.1193_{-0.0013}^{+0.0013}$ & $0.1167$ & $0.1218$\\ \hline 
$n_{s}$ & $0.9721_{-0.0043}^{+0.0042}$ & $0.9639$ & $0.9807$ & $0.9662_{-0.0047}^{+0.0045}$ & $0.9571$ & $0.9752$\\ \hline  
$10^{9}A_{s}$ & $2.063_{-0.032}^{+0.035}$ & $1.993$ & $2.133$ & $2.107_{-0.037}^{+0.033}$ & $2.037$ & $2.178$\\ \hline  
$\tau_{reio }$  & $0.0502_{-0.0082}^{+0.0092}$ & $0.0327$ & $0.0686$ & $0.0567_{-0.0092}^{+0.0077}$ & $0.0395$ & $0.0738$\\ \hline  
$w$ & $-0.9478_{-0.043}^{+0.022}$ & $-0.9999$ & $-0.8879$ & $-0.9808_{-0.018}^{+0.0042}$ & $-0.9999$ & $-0.9383$\\ \hline  
$100~\theta_{s }$  & $1.042_{-0.00031}^{+0.00030}$ & $1.042$ & $1.043$ & $1.042_{-0.00032}^{+0.00031}$ & $1.041$ & $1.043$ \\ \hline  \hline
$z_{reio }$ & $7.141_{-0.86}^{+0.89}$ & $5.298$ & $8.930$ & $7.883_{-0.84}^{+0.82}$ & $6.220$ & $9.597$\\ \hline  
$H_0$ \mbox{\tiny  $\frac{km}{sMpc}$ } & $67.88_{-0.96}^{+1.20}$ & $65.71$ & $69.97$ & $67.65_{-0.64}^{+0.80}$ & $66.14$ & $69.17$\\ \hline  
$\sigma_8$  & $0.7898_{-0.0093}^{+0.0120}$ & $0.7693$ & $0.8094$ & $0.7564_{-0.0110}^{+0.0120}$ & $0.7337$ & $0.7791$ \\ \hline  
$\Omega_{\rm m }$  & $0.3018_{-0.012}^{+0.0093}$ & $0.2817$ & $0.3226$& $0.3097_{-0.0087}^{+0.0081}$ & $0.2935$ & $0.3274$ \\ \hline\hline
${\log_{{10}}\,  \beta }$   & &  &    & $0.63_{-0.43}^{+0.40}$ & $-0.20$ & $1.49$ \\ \hline
\hline 
\end{tabular} 
\end{center}
\caption{In this table, we show the mean and $1\sigma$ values and the $2\sigma$ upper and lower limits for the cosmological and derived parameters for a $w$CDM model (left) and for the interacting model (right), where the prior is ${\log_{{10}}\,  \beta } \in [-8,4]$.}
\label{tab:fit}
\end{table}

\subsection{Flat prior on $\beta$}
Once we confirm the preferred scale of the coupling parameter is $\beta=\mathcal{O}(1-10)$, we can search for more precise constraints within this range by using flat priors for $\beta$ around the previous result. As we will see, the results are consistent with the ones obtained for the logarithmic prior so the significant improvement with respect to the non-interacting case cannot be attributed to a biased choice of the prior. Furthermore, the linear prior allows to reach the non-interacting case.

In Table~\ref{tab:fitlinear}, we show the  mean with $1\sigma$ confidence limits and the $2\sigma$ upper and lower limits for the parameter, again the $w$CDM model and the interacting model obtained using the linear prior. We confirm the previous results of a slight decrease in the baryon density $100~\Omega_{\rm b} h^2$ and an increase in the DM density $\Omega_{\rm dm}h^2$, thus a slightly higher value of the total matter density $\Omega_{\rm m}$. The biggest difference appears in the $\sigma_8$ parameter that has  a smaller value, which, as before, it is closely connected to the suppression of structures due to the interaction. Finally, we find, again, a $\Delta\chi^2=26$ improvement with respect to the $w$CDM model, without modifying the background cosmology. For all linear priors used, the results obtained are fully consistent with the ones of the logarithmic prior.

\begin{table}[!t]
\begin{center}
\renewcommand{\arraystretch}{1.8}
\begin{tabular}{ |c||c|c|c||c|c|c| } 
	\hline
	\hline
	\centering
	&\multicolumn{3}{c||}{$w$CDM model} &\multicolumn{3}{c|}{Elastic Interaction}\\ \hline
	Param. & mean$\pm\sigma$ & $2\sigma$ lower & $2\sigma$ upper  & mean$\pm\sigma$ & $2\sigma$ lower & $2\sigma$ upper\\
	\hline \hline \hline
$100\Omega_{\rm b } h^2$ & $2.264_{-0.015}^{+0.015}$ & $2.235$ & $2.294$ & $2.239_{-0.016}^{+0.015}$ & $2.208$ & $2.27$ \\\hline
$\Omega_{\rm dm}h^2$  & $0.1163_{-0.001}^{+0.001}$ & $0.1143$ & $0.1183$ & $0.1198_{-0.0011}^{+0.0012}$ & $0.1175$ & $0.1221$\\ \hline 
$n_{s}$ & $0.9721_{-0.0043}^{+0.0042}$ & $0.9639$ & $0.9807$ & $0.9649_{-0.0045}^{+0.0044}$ & $0.9559$ & $0.9738$\\ \hline  
$10^{9}A_{s}$ & $2.063_{-0.032}^{+0.035}$ & $1.993$ & $2.133$ & $2.114_{-0.036}^{+0.034}$ & $2.046$ & $2.187$\\ \hline  
$\tau_{reio }$  & $0.0502_{-0.0082}^{+0.0092}$ & $0.03266$ & $0.0686$ & $0.0575_{-0.0085}^{+0.0081}$ & $0.04119$ & $0.07465$\\ \hline  
$w$ & $-0.9478_{-0.043}^{+0.022}$ & $-0.9999$ & $-0.8879$ & $-0.9879_{-0.012}^{+0.0031}$ & $-0.9999$ & $-0.9507$\\ \hline  
$100~\theta_{s }$  & $1.042_{-0.00031}^{+0.0003}$ & $1.042$ & $1.043$ & $1.042_{-0.00032}^{+0.00032}$ & $1.041$ & $1.043$ \\ \hline  \hline
$z_{reio }$ & $7.141_{-0.86}^{+0.89}$ & $5.298$ & $8.93$ & $7.99_{-0.87}^{+0.77}$ & $6.278$ & $9.643$\\ \hline  
$H_0$ \mbox{\tiny  $\frac{km}{sMpc}$ } & $67.88_{-0.96}^{+1.2}$ & $65.71$ & $69.97$ & $67.64_{-0.58}^{+0.73}$ & $66.25$ & $69.03$\\ \hline  
$\sigma_8$  & $0.7898_{-0.0093}^{+0.012}$ & $0.7693$ & $0.8094$ & $0.751_{-0.010}^{+0.009}$ & $0.7318$ & $0.7701$ \\ \hline  
$\Omega_{\rm m }$  & $0.3018_{-0.012}^{+0.0093}$ & $0.2817$ & $0.3226$& $0.3109_{-0.0084}^{+0.0075}$ & $0.2948$ & $0.3272$ \\ \hline\hline
$\beta$   & &  &    & $11.6_{-10.8}^{+3.5}$ & $0.25$ & $26.79$ \\ \hline
\hline 
\end{tabular} 
\end{center}
\caption{In this table, we show the mean and $1\sigma$ values and the $2\sigma$ limits for the cosmological and derived parameters for a $w$CDM model (left) and for the interacting model (right). The results shown for the interacting model are obtained using the prior~$\beta\in[-0.03,30]$, but larger priors would lead to the same result with not well constrained $2\sigma$ regions and smaller priors would lead to smaller errors but less strong results.}
\label{tab:fitlinear}
\end{table}

\begin{figure}[!t]
	\centerline{
		\includegraphics[scale=0.47]{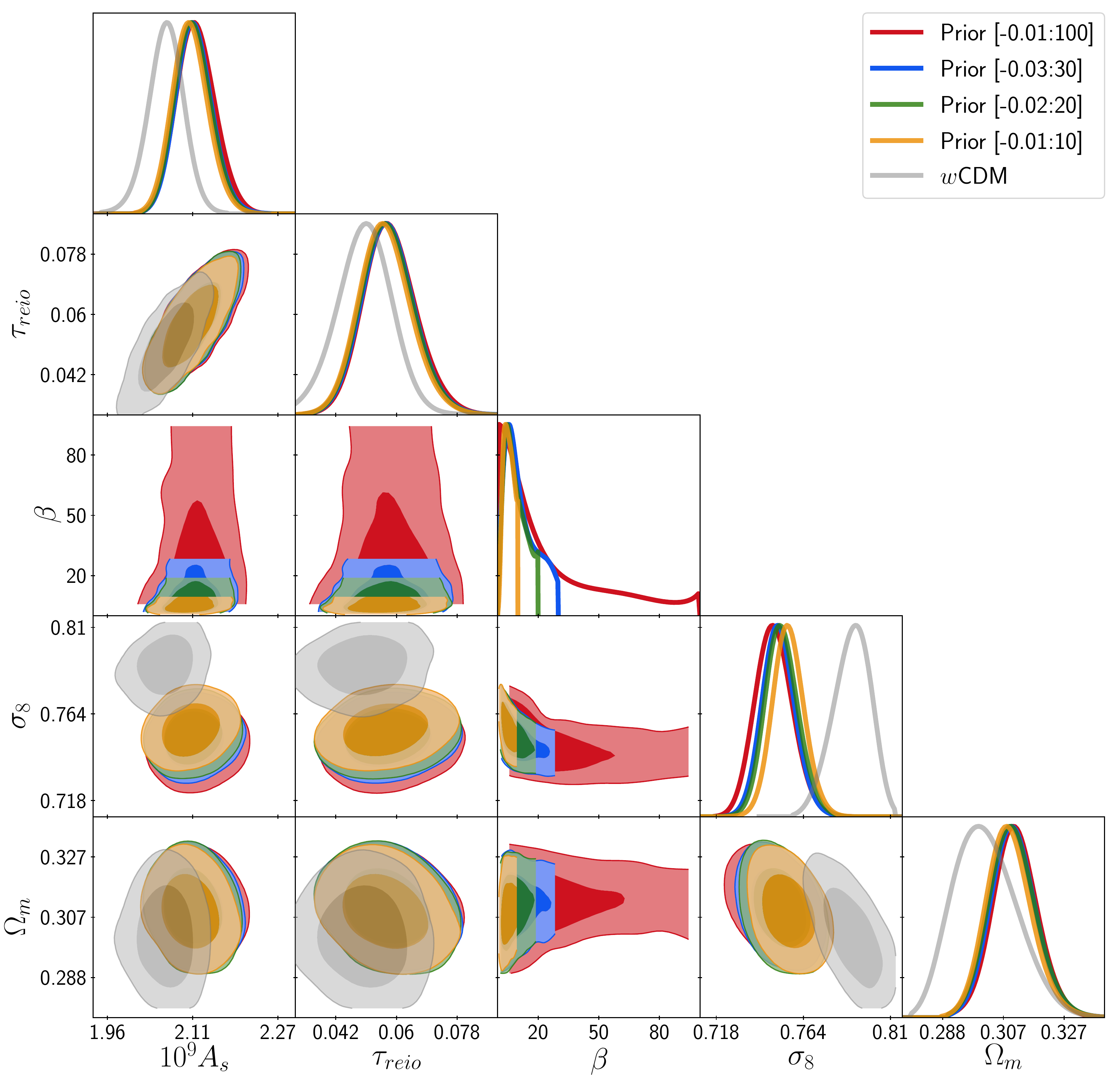}}
	\caption{The one-dimensional posterior  distributions and the two-dimensional contours obtained for several parameters using different priors on $\beta$, red line represents the prior $\beta\in[-0.01,100]$, blue line $\beta\in[-0.03,30]$, green line $\beta\in[-0.02,20]$ and yellow line $\beta\in[-0.01,10]$. The results presented before belong to the blue line, using the prior $\beta\in[-0.03,30]$. The gray lines are the reference model $w$CDM. As in the case of the logarithmic prior, we see the shift in $\sigma_8$ caused by the interaction and}
 	\label{Fig:prior}	
\end{figure}

The convergence of the chains is very poor when employing a flat linear prior over a sufficiently broad range for the coupling parameter $\beta$. The convergence is however improved when a reduced parameter range of the prior for $\beta$ is considered. For this reason we have explored different ranges for its prior as shown in Figure~\ref{Fig:prior}. We observe a modification in the confidence regions for $\beta$ as we change the prior, but only in what concerns their extension. It is clear that the posterior distributions are simply cut-off at the border of the prior. The remaining cosmological parameters posteriors are practically insensitive to the prior, proving the robustness of our constraints against the specific prior choice.\footnote{For $\sigma_8$ we do find a mild change due to the prior, as it is a derived parameter that strongly depends on the suppression of structures dictated by $\beta$. But such difference is negligible if we compare it to the change due to having the interaction or not having it, as we can infer from the one-dimensional posterior of $\sigma_8$ in Figure~\ref{Fig:prior}.}

In combination with the logarithmic prior we find data favour the interaction over the $w$CDM model having at more than $2\sigma$ a non vanishing value of the coupling parameter $\beta$, and also supported by the substantial improvement of the $\chi^2$ value. Although it may be premature to claim a detection of an interaction in view of our results, it is clear that they are encouraging to continue investigating this type of interactions with complementary data. For this purpose future data obtained from surveys like J-PAS~\cite{JPAS}, EUCLID~\cite{EUCLID} or DESI~\cite{DESI} would have a remarkable potential to confirm or rule out this interaction.

\subsection{Comparison to CMB polarisation from ACT and SPT}

\begin{figure}[!t]
	\centerline{\includegraphics[scale=0.3]{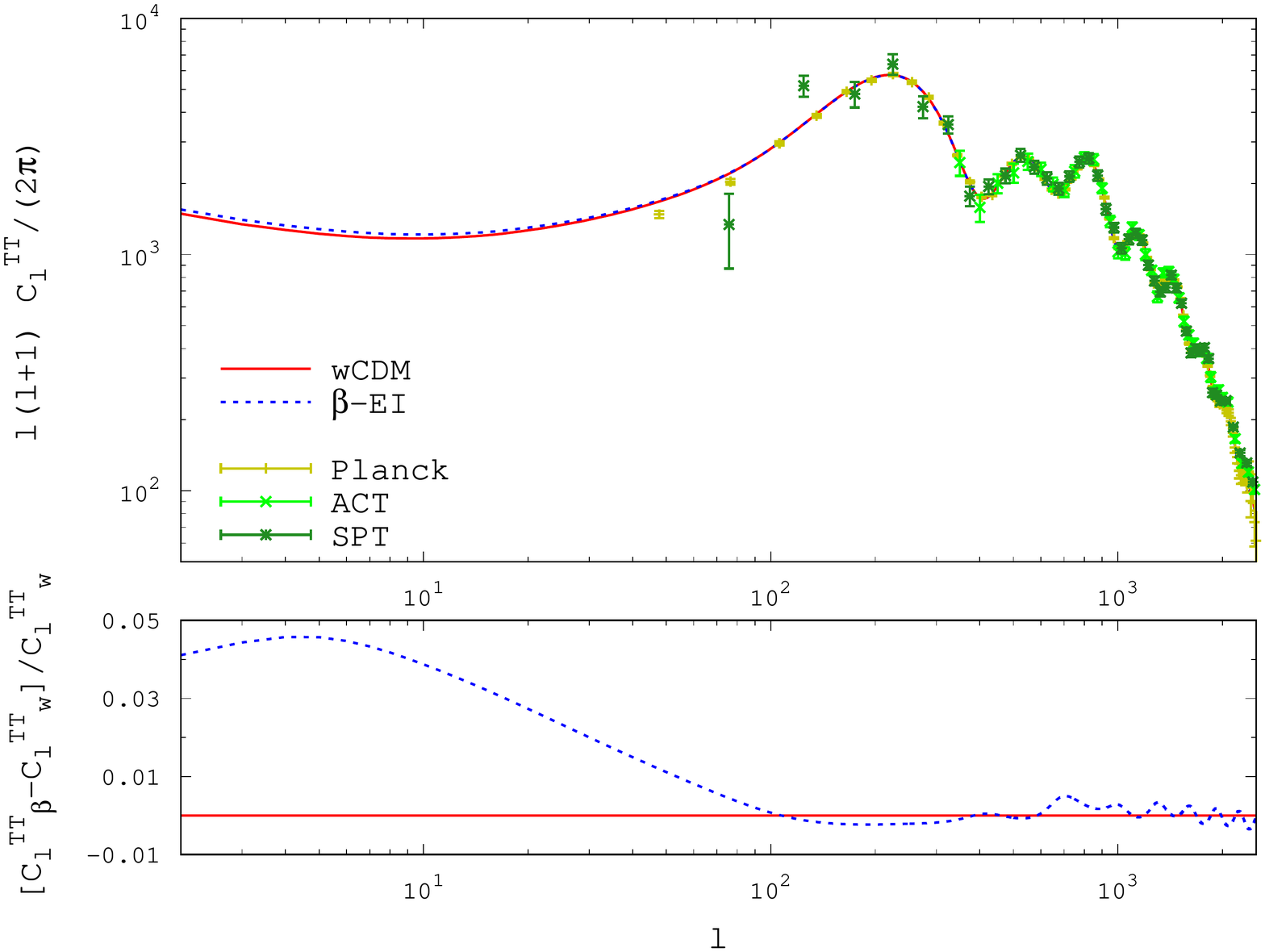}
	\includegraphics[scale=0.3]{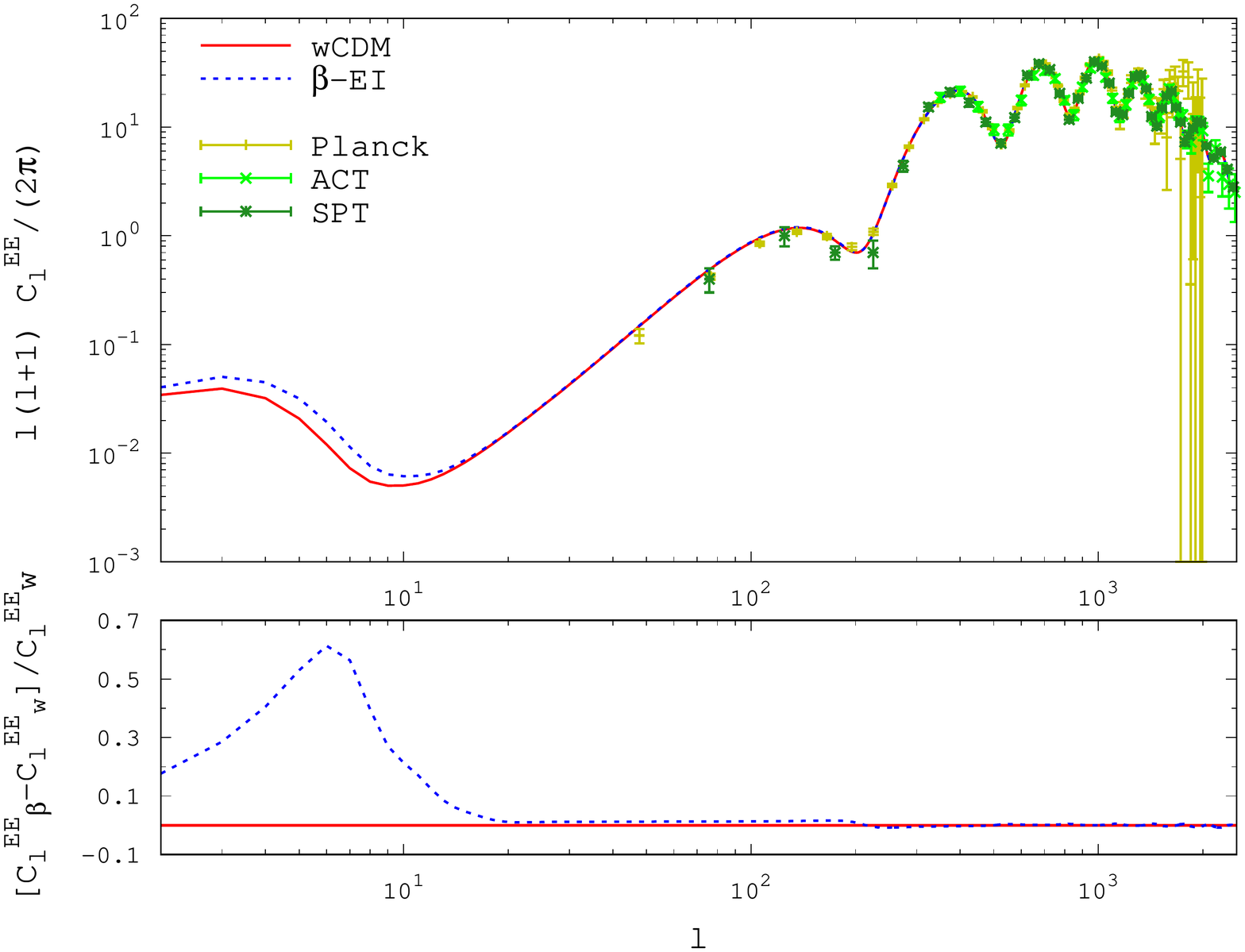}} 
	\centerline{\includegraphics[scale=0.3]{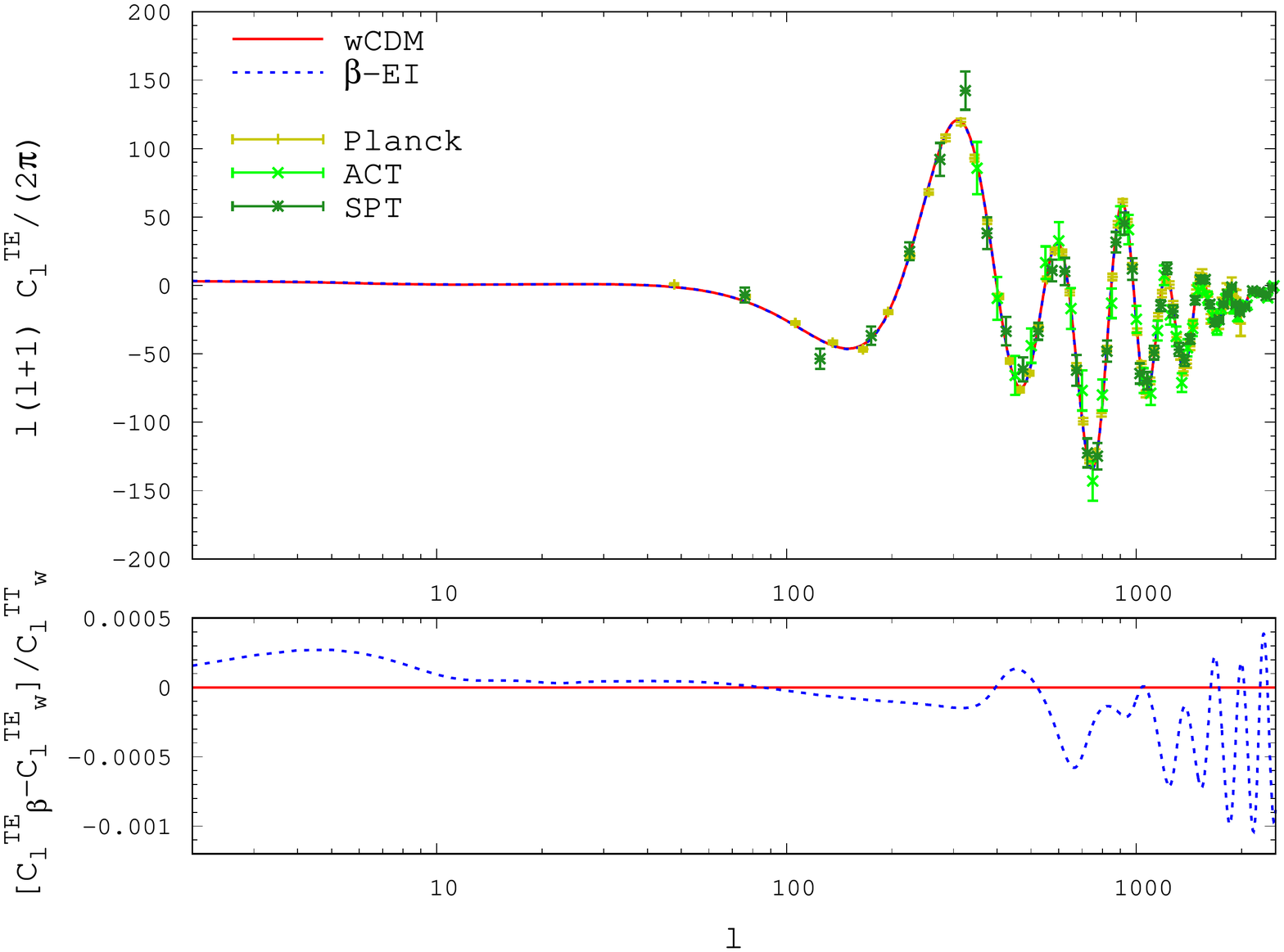}
	  \includegraphics[scale=0.3]{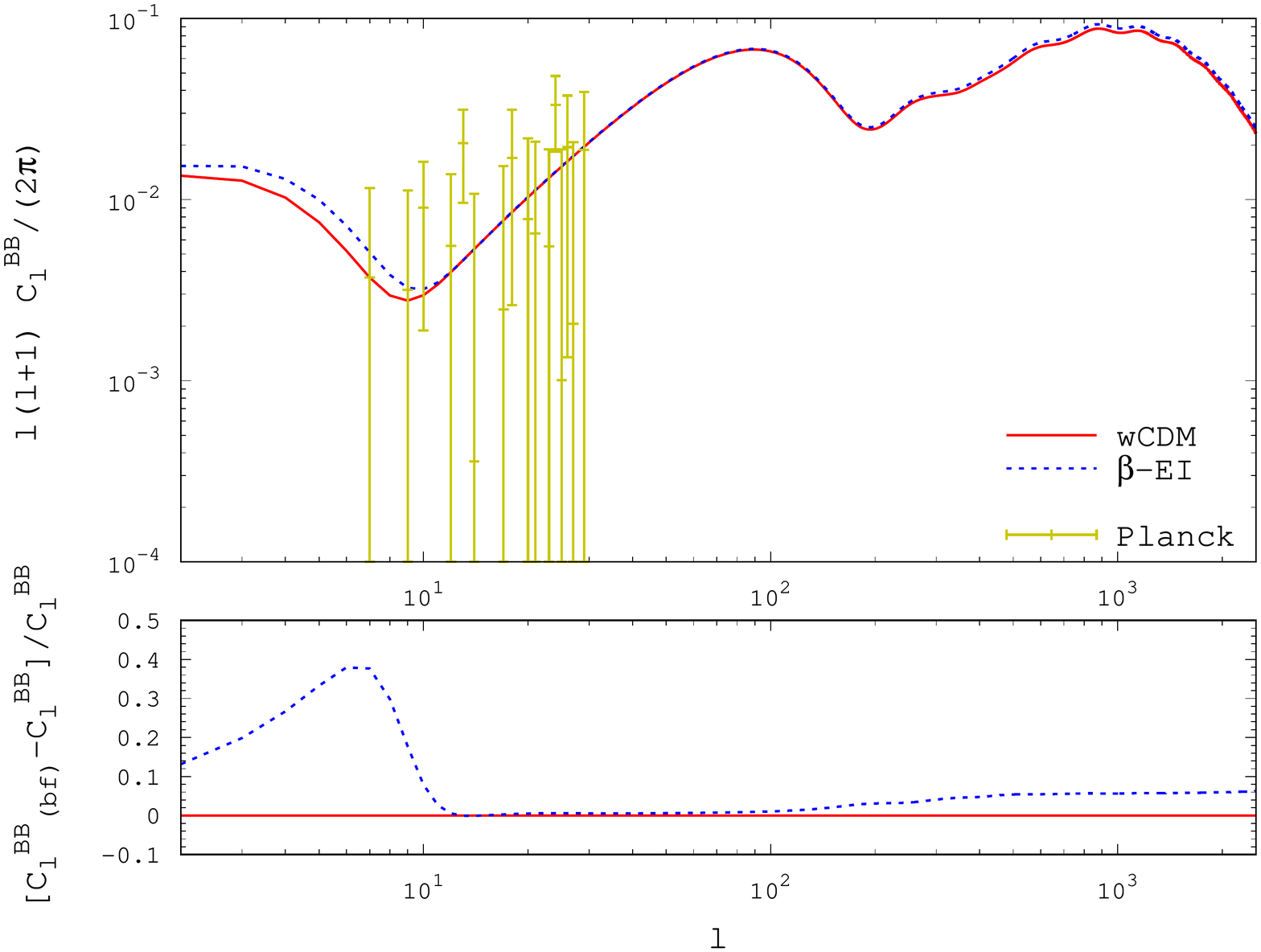}} 
	\caption{CMB angular power spectra and lensed potential for best-fit models presented in Table~\ref{tab:fitlinear} ($w$CDM and Elastic Interaction ($\beta$-EI)), together with temperature and polarisation data from Planck, ACT and SPT. }
	\label{fig:Cls_bf}	
\end{figure}

After obtaining the observational constraints, we will confront the polarisation spectra for the best fit parameters to current data from the Planck Collaboration \cite{Planck2019V}, the Atacama Cosmology Telescope (ACT) \cite{ACT2017} and the South Pole Telescope (SPT) \cite{SPTpol2018}. In Fig.~\ref{fig:Cls_bf} we can see that the model predictions are consistent with data, which should be seen as a confirmation of the goodness of the model. To be more quantitative, we compare the predictions of the best-fit parameters presented in Table~\ref{tab:fit} with the current available CMB temperature and polarisation data by computing the value of $\chi^2$ for the ACT and SPT polarisation data. We can see that the polarisation data from both ACT and SPT are compatible with the interacting model at the same level as $w$CDM with no significant deviation. In this respect, we should mention that we have used data up to $\ell =2500$, even though data is available up to $\ell=8300$ for ACT and $\ell=7500$ for SPT. The main limitation is the precision of the employed codes that does not allow us to reach those multipoles with sufficient precision.

\begin{table}[!t]
\begin{center}
\renewcommand{\arraystretch}{1.8}
\begin{tabular}{ |c||c|c|c||c|c|c| } 
	\hline
	\hline
	\centering
	&\multicolumn{3}{c||}{$w$CDM model} &\multicolumn{3}{c|}{Elastic Interaction}\\ 
\hline \hline \hline
 $\chi^2$ &  TT    &   EE   &   TE  &  TT    &   EE   &   TE \\ 
 \hline 
ACT    &  79.25  & 40.80   & 51.24   & 79.50  &  40.70  &  49.66 \\ \hline 
SPT    &  151.45  & 70.44    & 58.83 & 153.2  & 71.8   & 58.75
\\ \hline \hline 
\end{tabular} 
\end{center}
\caption{$\chi^2$-values using ACT and STP temperature and polarisation data for best-fit models presented in Table~\ref{tab:fit}. We see that the interacting model is reasonably consistent with these datasets at the same level as the $w$CDM model (except for the TT data of SPT).}
\label{tab:chi2} 
\end{table}


\section{Conclusions}
\label{sec:conclusions}

In this work, we have explored the possibility that DE could present a non-negligible interaction with ordinary baryonic matter on cosmological scales. We have employed a phenomenological model where only the Euler equations of baryons and DE are modified, while the continuity equations both at the background and perturbative level remain unchanged. The model only introduces one new parameter that governs the coupling and, unlike other interacting scenarios, the interaction becomes more relevant at late times thus enhancing the feasibility of having detectable observational signatures.
We have started by performing an analytical study of the modified perturbation equations in the different regimes attending to the relative hierarchy of the existing scales, namely: the Hubble expansion, the interaction rate, the effective sound horizon and the Fourier mode. Since the interaction rate is much smaller than the Hubble horizon at high redshift, the initial conditions for the cosmological perturbations are the same as in the standard model and are directly provided by the primordial spectrum of adiabatic perturbations. As the Universe expands, the interaction grows until it becomes the dominant contribution. The time at which the interaction affects the baryons and DE evolution differs due to the relative abundance of these components throughout the Universe evolution. The most significant observational signature that we have obtained from the interaction occurs for the sub-Hubble modes at late times of the baryons density contrast, which ceases its growth and saturates to a constant value. This results in a reduction of the small scales part of the total matter spectrum at low redshift, which in turn is in the correct direction to alleviate the $\sigma_8$ tension. We have also shown how the evolution of the peculiar velocities is modified which will result in interesting observational (potentially discriminating) signatures as it could modify the bias factor for galaxies as tracers of the underlying DM density field. It is important to notice the crucial difference with the case of DM-DE interactions considered in \cite{Asghari:2019qld}. In both cases, there is a modification of the peculiar velocities of the corresponding matter component that interacts with DE. However, we should bear in mind that galaxies can be considered virialised objects that follow the velocity field associated to the DM distribution. Thus, even if DM interacts with DE, the galaxies will still trace the DM velocity field. However, if baryons interact with DE, then they will be subject to an additional dragging by DE so they no longer provide perfect tracers of the DM density field, but an additional velocity bias is present. An obvious objection to this effect would be if the elastic interaction of baryons and DE remains the same on galactic scales, which may very well not be the case and for instance small scales viscosity effects or anisotropic stress could become relevant. These would however induce additional contributions to the bias.

We have confirmed our analytical findings with the numerical solutions obtained from modified versions of CLASS and CAMB that incorporate the effects of the interaction. The modified codes have then been used to obtain observational constraints. Due to the aforementioned ability of the interaction to suppress the growth of baryonic matter, the used datasets favour the presence of the interaction, with the non-interacting case at more than 2$\sigma$. For the analysis we have used two classes of priors on the interaction parameter. On one hand, we have imposed a flat prior on the logarithm of the parameter that has shown a preference for a non-vanishing interaction with $10^{-0.20}\leq\beta\leq 10^{1.49}$ at 2$\sigma$. We have also considered a flat prior on $\beta$, which has shown a poorer convergence than the logarithmic prior, but it provides 1$\sigma$ posterior distributions consistent with the results for the logarithmic prior, thus confirming the preference for the presence of the interaction.

The obtained results show the feasibility of obtaining cosmological traces of an interaction between DE and baryons and provide further motivation to seek for complementary observational tests that could discriminate such couplings. A promising route in this respect is cross-correlations between galaxy catalogues with the velocity field or CMB and CMB polarisation data. A distinctive feature of the interaction with baryons is the possibility of modifying the epoch of reionisation so a better understanding of this period will help discriminating the coupling of DE to baryons \cite{Koopmans:2015sua,Munoz:2019hjh}. On the other hand, intensity maps \cite{Bernal:2019jdo,Kovetz:2019uss}
can trace the distribution of Hydrogen gas so it will be a direct probe of the baryonic density field and, consequently, these data also provide a promising way of constraining the interactions considered in this work.

Finally, we have focused on a direct coupling of DE with baryons with a simple interaction as a proof-of-concept for the feasibility of having detectable signatures. However, it would be natural to expect interactions also with DM as in \cite{Asghari:2019qld} so it would be interesting to thoroughly find the degenerate directions in more general scenarios and analyse the orthogonality of different probes as possible ways of breaking them. In this respect, the suppression of the matter power spectrum induced by the interactions in these scenarios resembles the reduction of matter fluctuations on small effects due to the free streaming of massive neutrinos\footnote{We thank Bill Wright and Sunny Vagnozzi for pointing this out to us.} (see e.g. \cite{Hu:1997mj,Lesgourgues:2006nd,Lesgourgues:2014zoa,Cuesta:2015iho,Vagnozzi:2017ovm}). This suggests the presence of some degeneracy between neutrino masses and the interaction. However, the dependence of the background evolution on the neutrino masses but not on $\beta$, together with the slightly different $k$-dependence of the matter suppression may help breaking these degeneracies. On the other hand, some modified gravity or extended dark energy scenarios also predict a modification for the growth of structures and this could be compensated by the interaction, similarly to what happens with massive neutrinos \cite{Motohashi:2010sj,Baldi:2013iza,Hu:2014sea,Wright:2019qhf}. These are interesting issues that would be worth exploring in more detail.

\paragraph{Codes:}
Modified versions of the codes CLASS and CAMB for the computation of the evolution of linear perturbations are available on request.


\paragraph{Acknowledgments:} We would like to thank Miguel Aparicio Resco, Philippe Brax, Enea di Dio, Pierre Fleury, Antonio Maroto, David Mota, Sunny Vagnozzi and Bill Wright  for useful comments and discussions. The authors acknowledge support from the \textit{Atracci\'on  del Talento  Cient\'ifico} en Salamanca programme, from project PGC2018-096038-B-I00 by Spanish Ministerio de Ciencia, Innovaci\'on y Universidades and Ayudas del Programa XIII by USAL.

\appendix
\section{Equations in synchronous gauge}
\label{app:synch}
Since the interaction term is gauge invariant, i.e. under a transformation of the form $\delta x^\mu=\zeta^\mu$ it satisfies $\delta Q^\mu=-\mathcal{L}_\zeta Q^\mu=0$, at first order it is the same in any gauge. This implies that we only have modification in the velocity perturbation equations even if we make all of our computations in the synchronous gauge (in particular, in this case the temporal component of the 4-velocities of the fluid are not perturbed). 

 The perturbed line element for scalar modes in the synchronous gauge is given by
\begin{equation}
\dd s^2=a^2(\tau)^2 \left[-\dd \tau^2 + (\delta_{ij}+h_{ij}) \dd x^i \dd x^j\right]\;,
\end{equation}
where the perturbed spatial metric is written in terms of the scalar perturbations $h$ and $\eta$ as $h_{ij}={\rm  diag}(-2\eta,-2\eta,h+\eta)$ (without loss of generality, we assume that the perturbations propagate in the $z$-direction). Then, the conservation equations in the synchronous gauge read
\begin{eqnarray}
\deltab'&=&-\left(\thetab +\frac{1}{2}h'\right)\;,\\
\thetab'&=&-\mathcal{H} \thetab+ c_s^2k^2\deltab
+\Gamma_T(\theta_{\gamma}-\thetab)+\Gamma (\thetade-\thetab)\;,\\
\deltade'&=& -(1+w)\left(\thetade +\frac{1}{2}h'\right)-3 \mathcal{H} (\cs^2-w)\deltade- 9(1+w) \left(\cs^2-w\right) \frac{\mathcal{H}^2}{k^2}\thetade \;,\\
\thetade'&=&\left(-1+3\cs^2\right) \mathcal{H}\thetade +\frac{k^2 \cs^2}{1+w}\deltade-\Gamma R(\thetade-\thetab)\;.
\end{eqnarray}
where a constant DE equation of state parameter is assumed, and $\Gamma_T$, $\Gamma$ and $R\Gamma$ conserve the same previous definitions. 



\bibliography{biblio}
\bibliographystyle{utcaps}


\end{document}